\documentclass [a4paper, 10 pt]{article}

\usepackage{pstcol}
\usepackage{geometry}
\usepackage{amssymb}
\usepackage{relsize}
\usepackage{graphicx}
\usepackage{graphics}
\usepackage{pst-tree}
\usepackage{multicol}
\input xy

\newtheorem{Theorem}{Theorem}[section]
\newtheorem{Corollary}{Corollary}[section]
\newtheorem{Proposition}{Proposition}[section]

\newtheorem{Definition}{Definition}[section]

\newtheorem{Example}{Example}[section]

\newcommand{\blackrec}{\rule{1.25 mm}{2.5 mm}}

\newcommand*{\always}{\square}

\newcommand*{\nextt}{{\bigcirc}}

\newcommand*{\ang}[1]{\langle#1\rangle}

\begin{document}

\title{On relating CTL to Datalog}

\author {{\large Foto~Afrati$^{1}$} \  \  \  \
        {\large Theodore~Andronikos$^{1}$}  \ \  \  \
        {\large Vassia~Pavlaki$^{1}$ \thanks{The project  is co - funded by the European Social Fund (75\%) and National Resources (25\%)  - Operational Program for Educational and Vocational Training II (EPEAEK II) and particularly the Program HERAKLEITOS.}} \\
        {\large Eug\'enie Foustoucos$^{1}$}\\
        {\large Ir\`ene Guessarian$^{2}$}
\\
            {$^{1}$Dept. of Electrical and Computer Engineering}\\
            {National Technical University of Athens (NTUA)}\\
            {Athens, Greece}\\
            {\tt \{afrati,thandron,vpavlaki\}@softlab.ece.ntua.gr}\\
            {\tt eugenie@ermis.cs.ntua.gr} \\
            {$^{2}$Laboratory for Computer Science, LIAFA}\
            {Universit\'e Paris}\\
            {\tt ig@liafa.jussieu.fr}}

\maketitle

\begin{abstract}
    CTL is the dominant temporal specification language in practice mainly due to the fact that it admits model checking in linear time. Logic programming and the database query language Datalog are often used as an implementation platform for logic languages. In this paper we present the exact relation between CTL and Datalog and moreover we build on this relation and known efficient algorithms for CTL to obtain efficient algorithms for fragments of stratified Datalog. The contributions of this paper are:  a) We embed CTL into STD which is a proper fragment of stratified Datalog. Moreover we show that STD expresses exactly CTL -- we prove that by embedding STD into CTL. Both embeddings are linear. b) CTL can also be embedded to fragments of Datalog without negation. We define a fragment of Datalog with the successor build-in predicate that we call TDS and we embed CTL into TDS in linear time. We build on the above relations to answer open problems of stratified Datalog. We prove that query evaluation is linear and that containment and satisfiability problems are both decidable. The results presented in this paper are the first for fragments of stratified Datalog that are more general than those containing only unary EDBs.
\end{abstract}

\bibliographystyle{alpha}

\section{Introduction}

Temporal logics are modal logics used for the description and specification of the temporal ordering of events
\cite{Eme90}. Pnueli was the first to notice that temporal logics could be particularly useful for the specification
and verification of reactive systems \cite{Pnu77,Pnu81}. In defining temporal logics, there are two possible views
regarding the flow of time. One is that of linear time; at each moment there is only one possible future (Linear
Temporal Logic-LTL). The other is that of branching time (tree-like nature); at each moment time may follow different
paths which represent different possible futures \cite{EH86,Lam80}. The most prominent examples of the latter are CTL
(Computational Tree Logic), CTL$^\star$ (Full Branching Time Logic), and $\mu$-calculus.

Deciding whether a system meets a specification expressed in a language of temporal logic is called \textit{model
checking}. Model checking is decidable when the system is abstracted as a finite directed labeled graph and the
specification is expressed in a propositional temporal language. Model checking has been widely used for verifying the
correctness of, or finding design errors in many real-life systems \cite{CW96}. Through the 1990s, CTL has become the
dominant temporal specification language for industrial use \cite{Var01,CGL93} mostly due to its balance of expressive
power and linear model checking complexity. SMV \cite{McM93}, the first symbolic model checker (CTL­-based), and its
follower VIS \cite{BHS+96} (also CTL­-based), presented phenomenal success and serve as the basis for many industrial
model checkers.

The introduction of Datalog \cite{Ull88} represented a major breakthrough in the design of declarative, logic-oriented
database languages due to Datalog's ability to express recursive queries. Datalog is a rule-based language that has
simple and elegant semantics based on the notion of minimal model or least fixpoint. This leads to an operational
semantics that can be implemented efficiently, as demonstrated by a number of prototypes of deductive database systems
\cite{NT89,RSS92,ELMPS97}. Datalog queries are computed in polynomial time; however, it has been shown that Datalog
only captures a proper subset of monotonic polynomial-time queries \cite{ACY91}.

In order to express queries of practical interest, negation is allowed in the bodies of Datalog rules. Of particular
interest is stratified negation, which avoids the semantic and implementation problems connected with the unrestricted
use of nonmonotonic constructs in recursive definitions. In stratified Datalog \cite{ABW88,Ull88,CH85} negation is
allowed in any predicate under the constraint that negated predicates are computed in previous strata. Simple,
intuitive semantics leading to efficient implementation exists for stratified Datalog. Unfortunately, as shown in
\cite{Kol91}, this language has a limited expressive power as it can only express a proper subset of fixpoint queries.

We have three major contributions in this paper. The first contribution  is the definition of a fragment of stratified
Datalog (the class STD) which has the exact expressive power as CTL (Theorem \ref{expressivepower}). We prove that by
establishing a linear embedding from STD into CTL and vice versa. This is the first time that a fragment of stratified
Datalog is identified which  expresses exactly CTL. The definition of this fragment is simple and natural (see
Subsection \ref{STD}).

For our second contribution, we  build on the above result to solve open problems of stratified Datalog. More
specifically we prove that: a) query evaluation for STD is linear by reducing it to the model checking problem of CTL
and b) both satisfiability and containment problems are decidable for STD programs by reduction to the validity problem
of CTL. This is the first result that proves decidability of containment for a fragment of stratified Datalog which
uses EDB (Extensional Database) relations other than unary and hence it has not a limited number of nontrivial strata.
Checking containment of queries, i.e., verifying whether one query yields a subset of the result of the other, has been
the subject of research last decades. Query containment is crucial in many contexts such as query optimization, query
reformulation, knowledge-base verification, information integration, integrity checking and cooperative answering.
Table \ref{Table1} presents all known results on query containment for stratified Datalog including the results we
obtain here.

We also consider a fragment of a variant of Datalog without negation. We define the class TDS which is a fragment of
Datalog$_{Succ}$ and establish a linear embedding from CTL to TDS. Datalog$_{Succ}$ is Datalog enhanced with  the
build-in successor predicate and allows negation only in the EDB predicates. The successor predicate is needed to
express the universal quantifier which in stratified Datalog can be captured by using the full power of negation. Note
that  we use the conventional semantics  of Datalog and this constitutes a contribution relatively to previous works
\cite{GFAA03}. This is the third contribution.

\begin{table}
[h]
\begin{center}
\begin{tabular}
{||c||c||c||c||c||}
\hline
\hline
   &       &                            &                                  \\
     &   Stratified      &   STD                           &   Stratified negation                                 \\
    &   negation        &   (Stratified negation with unary      &   with unary                                           \\
    &                   &    + 1 binary EDB predicates)    &   EDB predicates                                      \\
        &                   &      &                                      \\
\hline
\hline
    Containment         &       undecidable                 &   EXPTIME--complete [Section 6]   &   decidable   \\
                &                                   &                           &   \cite{LMSS93,HMSS01}    \\
\hline
    Equivalence         &   undecidable                     &   EXPTIME--complete [Section 6]   &   decidable               \\
                        &                                   &                           &   \cite{LMSS93,HMSS01}    \\
\hline
    Satisfiability     &   undecidable                     &   EXPTIME--complete [Section 6]   &   decidable \\
     &                                   &                           &   \cite{LMSS93,HMSS01}    \\
\hline
    Evaluation          &   polynomial                      &   linear [Section 6]      &   linear [Section 6]     \\
\hline
\hline
\end{tabular}
\end{center}
\vspace{0.3cm}
\caption{Results on fragments of Stratified Datalog}\label{Table1}
\end{table}

\subsection{Motivating Examples}

The following three examples illustrate some of the subtle points of the translation of a CTL formula into stratified
Datalog and they are presented in order of increasing complexity. The subtleties in the case of Datalog$_{Succ}$ are of
similar nature. In all examples, we consider a  Kripke structure $\mathcal{K}$, which is given by: a set of states $W$,
the transition relation $R$ on the states, and atomic propositions assigned to the states.

\begin{Example}

This is the first motivating example for our translation techniques. Consider the CTL formula: $\varphi \equiv
\mathbf{E} \nextt p$. It says that, there exists a path starting from a state $s_0$ such that the next state on this
path is assigned the atomic proposition $p$. We may view the Kripke structure as a database $D$ with unary EDB
predicates for the atomic propositions $($here EDB predicate $P$ is associated to $p$$)$ and a binary EDB predicate $R$
for the transition relation. Now the following Datalog program says that if $s_0$ is computed in the  answers of the
query predicate $G$, then there exists a path in $D$ starting from $s_0$ which in one transition step reaches a state
where $P$ is true.

\begin{center}
{\footnotesize
$\left\{\begin{array}{l}
    G(x) \longleftarrow R(x,y), G_1(y)\\
    G_1(x) \longleftarrow P(x)\\
\end{array}\right. $
}
\end{center} \hfill$\blacktriangle$

\end{Example}

    Whereas this is not a recursive program, when the formula contains the ``until'' modality,  recursion is needed  as is the case in the example that follows.

\begin{Example}

Consider now the somewhat more complex formula $\varphi \equiv$ $\mathbf {E} \nextt p$ $\wedge$ $\mathbf{E}  (q
\mathbf{U} t)$. The Datalog query that expresses this formula is the following.

\begin{center}
{\footnotesize
$\left\{\begin{array}{l}
    G(x) \longleftarrow G_1(x), G_2 (x)    \\
    G_1(x) \longleftarrow R(x,y), G_3(y)  \\
    G_3(x) \longleftarrow P(x)  \\
    G_2(x) \longleftarrow G_4(x)   \\
    G_2(x) \longleftarrow G_5(x), R(x,y), G_2(y) \\
    G_4(x) \longleftarrow T(x) \\
    G_5(x) \longleftarrow Q(x) \\
\end{array}\right.$
}
\end{center}

This Datalog query expresses what the CTL formula says, i.e., there exists a path starting from a state $s_0$ that is
assigned $p$ on its next state and there is also a path $($different or the same$)$ such that it is assigned $q$ along
all its states up until it gets to a state that is assigned $t$. The second and third rules express the CTL formula
$\mathbf{E}\nextt p$, the four last rules express the CTL formula $\mathbf{E} (q \mathbf{U} t)$ and the first rule
asserts the conjunction of $\mathbf {E} \nextt p$ and $\mathbf{E}  (q \mathbf{U} t)$\footnote{It is easy to observe
that this particular Datalog program can be equivalently written using fewer rules. However we have written it here in
the form derived by our algorithm.}. \hfill$\blacktriangle$

\end{Example}

Now, there is a more complicated recursive case which requires a recursive predicate with two arguments and this is demonstrated in the third example.

\begin{Example} \label{example-intro}
Consider the CTL formula $\mathbf{E} \always p$ which can also be written as $\mathbf{E} (\bot \mathbf{\widetilde{U}} p)$. This formula says that there is an \textit{infinite} path from a state $s_0$ so that $p$ holds in all the states of the path. The existence of an infinite path on a finite Kripke structure is equivalent to the existence of a cycle. The following Datalog program expresses exactly this formula. In this program the rules with head predicate $W$ are ancillary, they just say that $x$ is an element of the domain $($the EDB predicates $P_i, 0 \leq i \leq n$, correspond to the atomic propositions$)$. They are used to obtain safe rules and to express true and false -- note that the second rule never fires, hence $G_2$ expresses  false and $G_1$ expresses true.  Thus the rules that express the essential meaning of the formula are 3--8.\\
\begin{center}
{\footnotesize
$\left\{\begin{array}{l}
                                                G_1(x) \longleftarrow W(x)\\
                                                G_2(x) \longleftarrow W(x), \neg G_1(x) \\
                                                G_3(x) \longleftarrow P(x)      \\
                                                G(x) \longleftarrow G_2(x), G_3(x)                                    \\
                                                G(x) \longleftarrow B(x,x)                                              \\
                                                G(x) \longleftarrow G_3(x), R(x,y), G(y)                                \\
                                                B(x,y) \longleftarrow G_3(x), R(x,y), G_3(y)                            \\
                                                B(x,y) \longleftarrow G_3(x), R(x,u), B(u,y)                            \\
                                                W(x)  \longleftarrow R(x,y)                                             \\
                                                W(x)  \longleftarrow R(y,x)                                             \\
                                                W(x)  \longleftarrow P_0(x)                                             \\
                                                \dots                                                                   \\
                                                W(x)  \longleftarrow P_n(x)                                             \\
                                                \end{array}\right. $
}
\end{center}

The two rules $($7th and 8th$)$ that compute $B$ $($combined with the third rule$)$ actually compute the transitive
closure  of $R$ over states where $P$ is true. The fifth rule says that the formula holds if there is a cycle starting
from  state $s_0$ with $P$ assigned to all its states. The sixth rule says that the formula holds if there is a path
which is followed by a cycle from  a state $s_0$ with $P$ assigned to all their states. \hfill $\blacktriangle$

\end{Example}

\subsection{Technical Challenges}

The examples illustrated the part of our contribution that translates a CTL formula to a Datalog query. However there
are a few technical challenges that do not show on these examples: 1) By a straightforward translation some Datalog
rules might not be safe (i.e., they may have variables that do not occur in nonnegated body subgoals). Thus we
introduce a number of rules which essentially define the domain by an IDB (Intentional Database) predicate which is
used in rules for safety -- this shows a little in Example \ref{example-intro}. 2) Trying to identify a fragment of
Datalog with exactly the same expressive power as CTL and use this fact to prove results for this fragment,  we have to
deal with the fact that CTL is interpreted over infinite paths. This means that finite Kripke structures over which we
interpret CTL have to be total on the binary relation $R$. Relational databases however over which Datalog programs are
interpreted do not have any constraints, i.e., the input could be any structure of the given schema. A solution to this
kind of problem that is suggested in the literature \cite{Eme90} is to add a self loop in those nodes that do not have
a successor in $R$. We adopt a similar solution only that we encompass it in the definitions of the Datalog fragment we
define, allowing thus for any input database to be captured. The example that follows explains further this point.

\begin{Example}
Consider the following Datalog program
\begin{center}
{\footnotesize
$\left\{\begin{array}{l}
A(x) \longleftarrow R(x,y) \\
G(x) \longleftarrow \neg A(x), P(x) \\
G(x) \longleftarrow R(x,y), P(y)
\end{array}\right. $
}
\end{center}

It is easy to see that it returns the same answer on any pair of databases which only differ in adding self loops in
$R$ on nodes that do not have a successor in $R$. \hfill $\blacktriangle$

\end{Example}

Finally, our results go through because CTL has the bounded model property which means that if there is a model for a
CTL formula then there is a finite model. Since in CTL infinite models are also assumed, in order to carry over results
to Datalog where finite input is assumed, we make use of this property.
\\
\\
The rest of the paper is organized as follows. Sections 2 and 3 are preliminary sections that define formally CTL
(Section 2),  Datalog,  Datalog$_{Succ}$  and stratified Datalog (Section 3). Section 4 presents the formalism of our
translation, discusses the notion of equivalence between CTL formulae and Datalog queries and defines the class of
Stratified Temporal Datalog (STD) programs which is a fragment of stratified Datalog. The embedding from CTL to STD is
also presented in Section 4. Section 5 gives the embedding from STD to CTL which is not straightforward so a discussion
on the technical challenges of this embedding is also included. In Section 6 we prove that query evaluation for STD
programs is linear and that checking containment and satisfiability is decidable. The embedding of CTL into
Datalog$_{Succ}$ is presented in Section 7. Finally, Section 8 shows how the present work can be extended to infinite
structures and discusses possible future research directions. The proof of Theorem \ref{Thrm: From CTL to TDS Soundness
& Completeness} is presented in the Appendix.

\subsection{Related Work}

Model checking is closely related to database query evaluation. The idea is based on the principle that Kripke
structures can be viewed as relational databases \cite{IV97}. One effective approach for efficiently implementing model
checking is based on the translation of temporal formulae into automata and has become an intensive research area
\cite{WVS83,VW86,VW94}. Another approach consists in translating temporal logics to Logic Programming  \cite{Llo87}.
Logic Programming  has been successfully used as an implementation platform for verification systems such as model
checkers. Translations of temporal logics such as CTL or $\mu$-calculus into logic programming can be found in
\cite{RRR+97,CDD+98,CP98}. \cite{CDD+98} presents the LMC project which uses XSB, a tabled logic programming system
that extends Prolog-style SLD resolution with tabled resolution.

The database query language Datalog has inspired work in \cite{GGV02}, where the language Datalog LITE is introduced.
Datalog LITE is a variant of Datalog that uses stratified negation, restricted variable occurrences and a limited form
of universal quantification in rule bodies. Datalog LITE is shown to encompass CTL and the alternation-free
$\mu$-calculus. Research  on model checking in the modal $\mu $-calculus is pursued in \cite{ZSS94} where the
connection between modal $\mu$-calculus and Datalog is observed. This is used to derive results about the parallel
computational complexity of this fragment of modal $\mu $-calculus.

In previous work \cite{GFAA03} we showed that the model checking problem for CTL can be reduced to the query evaluation
problem for fragments of Datalog. In more detail, \cite{GFAA03} presents a direct and modular translation from the
temporal logics CTL, ETL, FCTL (CTL extended with the ability to express fairness) and the modal $\mu$-calculus to
Monadic inf-Datalog with built-in predicates. It is called inf-Datalog because the semantics differ from the
conventional Datalog least fixed point semantics, in that some recursive rules (corresponding to least fixed points)
are allowed to unfold only finitely many times, whereas others (corresponding to greatest fixed points) are allowed to
unfold infinitely many times. The work in \cite{AAP+03}, which is a preliminary version of some of the results
presented here, embeds CTL into a fragment of Datalog$_{Succ}$.

We know that CTL can be embedded into Transitive Closure logic \cite{IV97} and into alternation-­free $\mu$--calculus
\cite{Eme96}. In \cite{GGV02} the authors observe that CTL can be embedded into stratified Datalog.  In this paper it
is the first time that the exact fragment of stratified Datalog with the same expressive power with CTL has been
identified.

Concerning containment of queries the majority of research refers to CQs. However there are important results
concerning also Datalog programs. In \cite{CGKV88} it was pointed out that query containment for monadic Datalog is
decidable. The work in \cite{Sag88} shows that checking containment of nonrecursive Datalog queries in Datalog queries
is decidable in exponential time. In \cite{CV97} it is shown that containment of Datalog queries in non-recursive
Datalog  is decidable in triply exponential time, whereas when the non-recursive query is represented as a union of
CQs, the complexity is doubly exponential. In \cite{LMSS93,HMSS01} authors proved that equivalence of stratified
Datalog programs is decidable but only for programs with unary EDB predicates. Our results are the first that encompass
also programs that contain  binary EDB predicates.

\section{CTL}

\subsection{Syntax and Semantics of CTL}

Temporal logics are classified as linear or branching according to the way they perceive the nature of time. In linear
temporal logics every moment has a unique future (successor), whereas in branching temporal logics every moment may
have more than one possible futures. Branching temporal logic formulae are interpreted over infinite trees or graphs
that can be unwound into infinite trees. Such a structure can be thought of as describing all the possible computations
of a nondeterministic program (branches stand for nondeterministic choices). Note that a  \textit{time step} is usually
identified with a computation step (e.g., a clock tick in a synchronous design). The future is considered to be the
reflexive future, it includes the present, and time is considered to unfold in discrete steps.

CTL (Computational Tree Logic) \cite{CE81,EC82} is a branching temporal logic that uses the path quantifiers
$\mathbf{E}$, meaning ``there exists a path", and  $\mathbf{A}$, meaning ``for all paths". A path is an infinite
sequence of states such that each state and its successor are related by the transition relation. The syntax of CTL
formulae uses temporal operators as well. For instance, to assert that ``property $\varphi$ is always true on every
path" or that ``there is a path on which property $\psi_1$ is true until $\psi_2$ becomes true" one writes $\mathbf{A}
\Box \varphi$ and $\mathbf{E} (\psi_1 \mathbf{U} \psi_2)$, respectively, where $\Box $ and $\mathbf{U}$ are temporal
operators. Various temporal operators are listed in the literature as part of the CTL syntax. However the operators
$\nextt$ and $\mathbf{U}$ form a complete set from which we can express all other operators. We give the syntax of CTL
in terms of these two temporal operators and later we also use the operator $\mathbf{\widetilde{U}}$, which facilitates
our translations.

    The syntax of CTL dictates that each usage of a temporal operator must be preceded by a path quantifier. These pairs consisting of the path quantifier and the temporal operator can be nested arbitrarily, but must have at their core a purely propositional formula. In the remaining of the paper $AP$ denotes the set of atomic propositions: $\{p_0, p_1, p_2, \ldots \}$ from which CTL formulae are built. We proceed to the formal definition of the syntax of CTL.\\
\\
S$_1$. \ Atomic propositions, $\top$ and $\bot$ are CTL formulae. \\
S$_2$. \ If $\varphi, \psi$ are CTL formulae then so are $\neg \varphi$, $\varphi \wedge \psi$, $\varphi \vee \psi$. \\
S$_3$. \ If $\varphi, \psi$ are CTL formulae then $\mathbf{E} \nextt \varphi$, $\mathbf{A} \nextt \varphi$, $\mathbf{E} (\varphi \mathbf{U} \psi)$, $\mathbf{A} (\varphi \mathbf{U} \psi)$ are CTL formulae. \\

The semantics of CTL is defined over temporal Kripke structures. A temporal Kripke structure $\mathcal{K}$ is a
directed labeled graph with node set $W$, arc set $R$ and labeling function $V$. $\mathcal{K}$ need not be a tree;
however, it can be turned into an infinite labeled tree if \textit{unwound} from a $s_0$ (see \cite{Eme90} and
\cite{Var97} for details). Below we give the definition of temporal Kripke structures.

\begin{Definition}  \label{Def: Kripke Structure}
    Let $AP$ be the set of atomic propositions. A \textit{temporal Kripke structure} $\mathcal{K}$ for $AP$ is a tuple $\langle W, R$, $V \rangle$, where:
\begin{itemize}
    \item   $W$ is the set of states,
    \item   $R \subseteq W \times W$ is the \emph{total} accessibility relation, and
    \item   $V: W \longrightarrow 2^{AP}$ is the valuation that determines which atomic propositions are true at each state.
\end{itemize}

    A \emph{finite} Kripke structure $\mathcal{K}$ is a Kripke structure $\ang{W, R, V}$ with finite $W$. \hfill\blackrec
\end{Definition}

    In Kripke structures the set of states $W$ can be \textit{infinite}. $W$ as defined in Definition \ref{Def: Kripke Structure} may be of any cardinality. In this paper we are interested in relational databases, where the universe $W$ is \textit{finite}. Hence, our Kripke structures are finite. In CTL we are dealing with \textit{infinite} computation paths, which means that in order for the accessibility relation $R$ to be meaningful, $R$ must be \textit{total} (\cite{KVW00}):

\begin{equation}    \label{Eq: The Totality of R}
    \forall x\exists y \ R(x, y)
\end{equation}

\begin{Definition}  \label{Def: Path}
    A \textit{path} $\pi$ of $\mathcal{K}$ is an infinite sequence $s_0, s_1, s_2,\ldots$ of states of $W$, such that $R(s_i, s_{i+1})$, $i\geq 0$. We also use the notational convention $\pi^i = s_i, s_{i+1}, s_{i+2}, \ldots$ .  \  \hfill\blackrec
\end{Definition}

     The notation $\mathcal{K}, s \models \varphi$ means that ``the formula $\varphi$ holds at state $s$ of $\mathcal{K}$''. The meaning of $\models$ is formally defined as follows:

\begin{Definition}  \label{Def: Models} \
\begin{itemize}
        \item   $\models \top$ and \ $\not \models \bot$
        \item   $\mathcal{K}, s \models p \Longleftrightarrow p \in V(s)$, for an atomic proposition $p \in AP$
        \item   $\mathcal{K}, s \models \neg \varphi \Longleftrightarrow \mathcal{K}, s \not \models \varphi$
        \item   $\mathcal{K}, s \models \varphi \vee \psi\Longleftrightarrow \mathcal{K}, s \models \varphi$ or $\mathcal{K}, s \models \psi$
        \item   $\mathcal{K}, s \models \varphi \wedge \psi \Longleftrightarrow \mathcal{K}, s \models \varphi$ and $\mathcal{K}, s \models \psi$
        \item   $\mathcal{K}, s \models \mathbf{E} \varphi \Longleftrightarrow$ \emph{there exists a path} $\pi = s_0, s_1, \ldots$, \emph{with initial state} $s = s_0$, \emph{such that} $\mathcal{K}, \pi \models \varphi$
        \item   $\mathcal{K}, s \models \mathbf{A} \varphi \Longleftrightarrow$ \emph{for every path} $\pi = s_0, s_1, \ldots$, \emph{with initial state} $s = s_0$ \emph{it holds that} $\mathcal{K}, \pi \models \varphi$
          \item   $\mathcal{K}, \pi \models \nextt \varphi \Longleftrightarrow \mathcal{K},\pi^1 \models \varphi$
        \item   $\mathcal{K}, \pi \models \varphi \mathbf{U} \psi \Longleftrightarrow$ \emph{there exists} $i \geq 0$ \emph{such that} $\mathcal{K}, \pi^i \models \psi$ \emph{and for all} $j,\ 0\leq j < i,$ $\mathcal{K}, \pi^j \models \varphi$
        \item   $\mathcal{K}, \pi \models \varphi \mathbf{\widetilde{U}} \psi \Longleftrightarrow$ \emph{for all}
$i\geq 0$ \emph{such that} $\mathcal{K}, \pi^i \not \models \psi$ \emph{there exists} $j,\ 0\leq j<i,$ \emph{such that} $\mathcal{K}, \pi^j \models \varphi$ \hfill\blackrec
\end{itemize}
\end{Definition}

A CTL state formula $\varphi$ is \emph{satisfiable} if there exists a Kripke structure $\mathcal{K} = \ang{W, R, V}$
such that $\mathcal{K}, s \models \varphi$, for some $s \in W$. In this case $\mathcal{K}$ is a \emph{model} of
$\varphi$. If $\mathcal{K}, s \models \varphi$ for every $s \in W$, then $\varphi$ is \emph{true} in $\mathcal{K}$,
denoted $\mathcal{K} \models \varphi$. If $\mathcal{K} \models \varphi$ for every $\mathcal{K}$, then $\varphi$ is
\emph{valid}, denoted $\models \varphi$. If $\mathcal{K} \models \varphi$ for every finite $\mathcal{K}$, we say that
$\varphi$ is valid with respect to the class of finite Kripke structures, denoted $\models_f \varphi$.

The \emph{truth set} of a CTL formula $\varphi$ with respect to a Kripke structure $\mathcal{K}$ is the set of states
of $\mathcal{K}$ at which $\varphi$ is true. We define formally the truth set as follows:

\begin{Definition} \label{Def: Truth Set} $($Truth set$)$
    Given a CTL formula $\varphi$ and a Kripke structure $\mathcal{K} = \ang{W, R, V}$, the \emph{truth set} of $\varphi$ with respect to $\mathcal{K}$, denoted $\varphi[\mathcal{K}]$, is $\{s \in W \ | \ \mathcal{K}, s \models \varphi \}$.   \hfill \blackrec
\end{Definition}

\subsection{Normal Forms}

    CTL formulae can be transformed in two normal forms: \textit{existential normal form} and \textit{positive normal form}. The translations we give in Sections \ref{From CTL to STD} and \ref{From CTL to TDS} cover each of these two syntactic variations of CTL.

\subsubsection{Existential Normal Form}

In \textit{existential normal form} negation is allowed to appear in front of CTL formulae. The universal path quantifier $\mathbf{A}$ is cast in terms of its dual existential path quantifier $\mathbf{E}$ using negation and the temporal operator $\mathbf{\widetilde{U}}$: $\mathbf{A} (\psi_1 \mathbf{U} \psi_2)$ becomes $\neg \mathbf{E} (\neg \psi_1 \mathbf{\widetilde{U}} \neg \psi_2)$. The $\mathbf{\widetilde{U}}$ operator was initially introduced in \cite{Var98,KVW00} as the dual operator of $\mathbf{U}$. One can think of $\mathbf{E} (\psi_1 \mathbf{\widetilde{U}} \psi_2)$ as saying that there exists a path on which:\\
    (1)     either $\psi_2$ always holds, or \\
    (2)     the first occurrence of $\neg \psi_2$ is strictly preceded by an occurrence of $\psi_1$.

    In general, every CTL formula can be written in \textit{existential normal form} using negation, the temporal operators $\nextt$, $\mathbf{U}$, $\mathbf{\widetilde{U}}$ and the existential path quantifier $\mathbf{E}$ (without the universal path quantifier $\mathbf{A}$). The syntax in this case is given by rules S$_1'$-S$_3'$ and Proposition \ref{Prop: ENF} states formally the equivalence of the two forms. \\
\\
S$_1'$. \ Atomic propositions and $\top$ are CTL formulae. \\
S$_2'$. \ If $\varphi, \psi$ are CTL formulae then so are $\neg \varphi$, $\varphi \wedge \psi$. \\
S$_3'$. \ If $\varphi, \psi$ are CTL formulae then $\mathbf{E} \nextt \varphi$, $\mathbf{E} (\varphi \mathbf{U} \psi)$ and $\mathbf{E} (\varphi \mathbf{\widetilde{U}} \psi)$ are CTL formulae.

\begin{Proposition} \label{Prop: ENF}
    Every CTL formula $\varphi$ can be transformed into a CTL formula $\varphi'$ in \textit{existential normal form} such that $\mathcal{K}, s \models \varphi$ \ iff  \ $\mathcal{K}, s \models \varphi'$ for every $\mathcal{K} = \ang{W, R, V}$ and every $s \in W$.
\end{Proposition}

\noindent\textbf{Proof}\\
    The universal path quantifier $\mathbf{A}$ is expressed as follows: $\mathbf{A} \nextt \psi$  is rewritten as  $\neg \mathbf{E} \nextt \neg \psi$, $\mathbf{A} (\psi_1  \mathbf{U} \psi_2)$ as $\neg \mathbf{E} (\neg \psi_1 \mathbf{\widetilde{U}} \neg \psi_2)$ and $\mathbf{A} (\psi_1  \mathbf{\widetilde{U}} \psi_2)$ as $\neg \mathbf{E} (\neg \psi_1 \mathbf{U} \neg \psi_2)$. The correctness of these transformations follows immediately from Definition \ref{Def: Models}. Also $\bot$ can be viewed as an abbreviation of $\neg \top$. \hfill$\dashv$ \\

The translation presented in Section \ref{From CTL to STD} translates CTL formulae in existential normal form into
stratified Datalog. As the universal quantifier is not used, stratified Datalog expresses nicely CTL formulae.

\subsubsection{Positive Normal Form \cite{Var98}}

    Every CTL formula can be equivalently written in \textit{positive normal form} where negation is applied only on atomic propositions. However, to compensate for the loss of full negation we need to use also the temporal operator $\mathbf{\widetilde{U}}$. Every CTL formula can be written in positive normal form using negation applied only on atomic propositions, the temporal operators $\nextt$, $\mathbf{U}$ and $\mathbf{\widetilde{U}}$ and both existential $\mathbf{E}$ and universal $\mathbf{A}$ path quantifiers. This is achieved by pushing negations inward as far as possible using De Morgan's laws and dualities of path quantifiers and temporal operators. The syntax of CTL in this case is given by rules S$_1''$-S$_3''$. \\
\\
S$_1''$. \ Atomic propositions, $\top$ and their negation are CTL formulae. \\
S$_2''$. \ If $\varphi, \psi$ are CTL formulae then so are $\varphi \wedge \psi$, $\varphi \vee \psi$. \\
S$_3''$. \ If $\varphi, \psi$ are CTL formulae then $\mathbf{E} \nextt \varphi$, $\mathbf{A} \nextt \varphi$, $\mathbf{E} (\varphi \mathbf{U} \psi)$, $\mathbf{A} (\varphi \mathbf{U} \psi)$, $\mathbf{E} (\varphi \mathbf{\widetilde{U}} \psi)$ and $\mathbf{A} (\varphi \mathbf{\widetilde{U}} \psi)$ are CTL formulae. \\

\begin{Proposition} \label{Prop: PNF}
    Every CTL formula $\varphi$ can be transformed into a CTL formula $\varphi'$ in \textit{positive normal form} such that $\mathcal{K}, s \models \varphi$ \ iff  \ $\mathcal{K}, s \models \varphi'$ for every $\mathcal{K} = \ang{W, R, V}$ and every $s \in W$.
\end{Proposition}

\noindent\textbf{Proof}\\
    The proof can be found in \cite{Var98}.\hfill $\dashv$ \\

The translation in Section \ref{From CTL to TDS} considers CTL formulae in positive normal form and translates them
into Datalog enhanced with the $Succ$ operator; the latter is needed to express the universal path quantifier. It turns
out that in this translation there is no need for negation in recursively defined predicates. Table \ref{Tbl: Normal
Forms} presents the two normal forms in which a CTL formula can be written in, and the corresponding fragments of
Datalog used for the translation.

\begin{table}  [h]
\begin{center}
\begin{tabular}{|c|c|c|c|}
    \hline
      Translation & CTL Normal Form & Datalog \\
    &   &   \\
    \hline
    \hline
    [Section \ref{From CTL to STD}] & Existential Normal Form & Stratified Datalog   \\
    \hline
    [Section \ref{From CTL to TDS}] & Positive Normal Form & Datalog + $Succ$       \\
    \hline
\end{tabular}
\vspace{0.5cm}
\caption{Normal forms vs. Datalog fragments} \label{Tbl: Normal Forms}
\end{center}
\end{table}

\subsection{Model Checking and Complexity}

\emph{Model checking} is the problem of verifying the conformance of a finite state system to a certain behavior, i.e.,
verifying that the labeled transition graph satisfies (is a model of) the formula that specifies the behavior.
 Hence, given a labeled transition graph $\mathcal{K}$, a state $s$ and a temporal formula $\varphi$, the model checking problem for $\mathcal{K}$ and $\varphi$ is to
decide whether $\mathcal{K},s \models \varphi$. The \emph{size} of the labeled transition system $\mathcal{K}$, denoted
$|\mathcal{K}|$, is taken to be $|W| + |R|$ and the \emph{size} of the formula $\varphi$, denoted $|\varphi|$, is the
number of symbols in $\varphi$.

For CTL formulae the model checking problem is known to be P--hard \cite{Schno03}, something that makes highly
improbable the development of efficient parallel algorithms. However, there exist efficient algorithms that solve it in
$O(|\mathcal{K}| |\varphi|)$ time \cite{CES86}. It is insightful to examine how the two parameters $|\mathcal{K}|$ and
$|\varphi|$ affect the complexity. This can be done by introducing the following two complexity measures for the model
checking problem \cite{VW86}:

\begin{itemize}
    \item   \textit{data}  complexity, which assumes a \textit{fixed} formula and \textit{variable} Kripke structures, and
    \item   \textit{program} or \textit{formula} complexity, which refers to \textit{variable} formulae over a \textit{fixed} Kripke structure.
\end{itemize}

CTL model checking is NLOGSPACE--complete with respect to data complexity\footnote{In real life examples the crucial
factor is $|\mathcal{K}|$, which is much larger than $|\varphi|$.} and its formula complexity is in $O(\log|\varphi|)$
space \cite{Schno03}. Another important problem for CTL is the \emph{validity} problem, that is deciding whether a
formula $\varphi$ is valid or not. This problem is much harder; it has been shown to be EXPTIME­-complete \cite{Var97}.
The following two theorems state known results of CTL on which we built in Section \ref{containment} to argue about
stratified Datalog.

\begin{Theorem}     \label{Thrm: CTL Validity} $($Validity$)$ $\cite{Var97}$    The validity problem for CTL is EXPTIME--complete.
\end{Theorem}

    CTL exhibits another important property, namely the bounded model property: if a formula $\varphi$ is satisfiable, then $\varphi$ is satisfiable in a structure of bounded cardinality.\footnote{As M. Vardi remarks in \cite{Var97} this is stronger than the finite model property which says that if $\varphi$ is satisfiable, then $\varphi$ is satisfiable in a finite structure.}

\begin{Theorem}     \label{Thrm: CTL BMP} $($Bounded Model Property$)$ $\cite{Eme90}$   If a CTL formula $\varphi$ has a model, then $\varphi$ has a model with at most $2^{|\varphi|}$ states.
\end{Theorem}

\section{Datalog}

Datalog \cite{Ull88} is a query language for relational databases. An \emph{atom} is an expression of the form $E(x_1,
\ldots, x_r)$, where $E$ is a predicate symbol and $x_1, \ldots, x_r$ are either variables or constants. A
\textit{ground fact} (or \emph{ground atom}) is an atom of the form $E(c_1, \ldots, c_r)$, where $c_1, \ldots, c_r$ are
constants. From a logic perspective, a relation $\widehat{E}$ corresponding to predicate symbol $E$ is just a finite
set of ground facts of $E$ and a relational database $D$ is a finite collection of relations. To simplify notation, in
the rest of this paper we use the same symbol for the relation and the predicate symbol; which one is meant will be
made clear by the context.

\begin{Definition}  \emph{\cite{DEG+01}}
    A \emph{database schema} $\mathfrak{D}$ is an ordered tuple $\langle W, E_1$, \ldots, $E_n \rangle$, where $W$ is the domain of the schema and $E_1, \ldots, E_n$ are predicate symbols, each with its associated arity.

    Given a database schema $\mathfrak{D}$, the set of all ground facts formed from $E_1, \ldots, E_n$ using as constants the elements of $W$ is denoted $\mathcal{H}_B (W)$. A \emph{database} $D$ over $\mathfrak{D}$ is a \emph{finite} subset of $\mathcal{H}_B (W)$; in this case, we say that $\mathfrak{D}$ is the underlying schema of $D$. The \emph{size} of a database $D$, denoted $|D|$, is the number of ground facts in $D$.   \hfill \blackrec
\end{Definition}

\begin{Definition}  A \emph{Datalog program} $\Pi$ is a finite set of function-free Horn clauses, called \emph{rules}, of the form:
\begin{center}
$    G(x_1,\ldots,x_n)\longleftarrow B_1(y_{1,1}, \ldots, y_{1,n_1}), \ldots, B_k(y_{k,1}, \ldots, y_{k,n_k})$ \end{center}
    where:  \\
   -   $x_1,\ldots,x_n$ are variables, \\
   -   $y_{i,j}$'s are either variables or constants, \\
   -   $G(x_1,\ldots,x_n)$ is a predicate atom, called the \emph{head} of the rule, and  \\
   -   $ B_1(y_{1,1}, \ldots, y_{1,n_1})$, \ldots, $B_k(y_{k,1}, \ldots, y_{k,n_k})$ are atoms that comprise the \emph{body} of the rule.   \hfill \blackrec
\end{Definition}

Predicates that appear in the head of some rule are called \textit{IDB} (Intensional Database) predicates , while
predicates that appear only in the bodies of the rules are called \textit{EDB} (Extensional Database) predicates. Each
Datalog program $\Pi$ is associated with an ordered pair of database schemas $(\mathfrak{D}_i, \mathfrak{D}_o)$, called
the \emph{input-output schema}, as follows: $\mathfrak{D}_i$ and $\mathfrak{D}_o$ have the \textit{same} domain and
contain exactly the EDB and IDB predicates of $\Pi$, respectively. Given a database $D$ over $\mathfrak{D}_i$ the set
of ground facts for the IDB predicates, which can be deduced from $D$ by applications of the rules in $\Pi$, is the
output database $D'$ (over $\mathfrak{D}_o$), denoted $\Pi(D)$. Databases over $\mathfrak{D}_i$ are mapped to databases
over $\mathfrak{D}_o$ via $\Pi$.

\begin{Definition}
    Given a Datalog program $\Pi$ we distinguish an IDB predicate $G$ and call it the \emph{goal} $($or \emph{query}$)$ predicate of $\Pi$. Let $D$ be an input database\footnote{In the sequel of the paper we assume without explicitly mentioning it, that the input databases for a Datalog program $\Pi$ have the appropriate schema.}; The \emph{query evaluation} problem for $G$ and $D$ is to compute the set of ground facts of $G$ in $\Pi(D)$, denoted $G_{\Pi}(D)$. \hfill \blackrec
\end{Definition}

    The \emph{dependency graph} of a Datalog program is a directed graph with nodes the set of IDB predicates of the program; there is an arc from predicate $B$ to predicate $G$ if there is a rule with head an instance of $G$ and at least one occurrence of $B$ in its body. The \emph{size} of a rule $r$, denoted $|r|$, is the number of symbols appearing in $r$. Given a Datalog program
$\Pi =  \left\{\begin{array}{l}
        r_n         \\
        \ldots      \\
        r_0         \\
        \end{array}\right.$,
the \emph{size} of $\Pi$, denoted $|\Pi|$, is $|r_0| + \ldots + |r_n|$.
\\
\\
\noindent \textbf{Stratified Datalog}
\\
\\
\noindent Intuitively, stratified Datalog is a fragment of Datalog with negation allowed in any predicate under the
constraint that negated predicates are computed in previous strata. Each head predicate of $\Pi$ is a head predicate in
precisely one stratum $\Pi_i$ and appears only in the body of rules of higher strata $\Pi_j$ ($j>i$) \cite{GGV02}. In
particular this means that:

\begin{enumerate}
    \item   If $G$ is the head predicate of a rule that contains a negated $B$ as a subgoal, then $B$ is in a lower stratum than $G$.
    \item   If $G$ is the head predicate of a rule that contains a non negated $B$ as a subgoal, then the stratum of $G$ is at least as high as the stratum of $B$.
\end{enumerate}

In other words a program $\Pi$ is stratified, if there is an assignment $str()$ of integers $0, 1, \ldots$ to the predicates in $\Pi$, such that for each clause $r$ of $\Pi$ the following holds: if $G$ is the head predicate of $r$ and $B$ a predicate in the body of $r$, then $str(G) \geq str(B)$ if $B$ is non negated, and $str(G) > str(B)$ if $B$ is negated.

\begin{Example} For the stratified program:
   \begin{center}
{\small $\left\{\begin{array}{l}
        A \longleftarrow \neg B  \\
        B \longleftarrow \neg C    \\
        C \longleftarrow  D  \\
    \end{array}\right.$ }
    \end{center}

\noindent $str()$ is the following: $str(C) = str(D) = 0, str(B) = 1$ and $str(A) = 2$. \hfill $\blacktriangle$
\end{Example}

The dependency graph can be used to define strata in a given program. In the dependency graph of a stratified program
$\Pi$, whenever there is a rule with head predicate $G$ and negated subgoal predicate $B$, there is no path from $G$ to
$B$. That is there is no recursion through negation in the dependency graph of a stratified program. The number of
strata of $\Pi$ is denoted $strata(\Pi)$. For more details on stratified Datalog see \cite{Ull88,ZCF+97}.
\\
\\
\noindent \textbf{Datalog$_{Succ}$}
\\
\\
\noindent Datalog$_{Succ}$ is Datalog where the domain is totally ordered and which uses the binary build-in predicate
$Succ(X,Y)$ to express that $Y$ is the successor of $X$, where $X$ and $Y$ take values from a totally ordered domain.
Papadimitriou in \cite{Pap85} proved that Datalog$_{Succ}$ captures polynomial time.

Notice that the term  ``successor'' is overloaded in the following sense. In the literature on CTL successor is used to
refer to the second argument of $R(x,y)$ and we say that $y$ is the child of $x$. In Datalog$_{Succ}$ the build-in
predicate $Succ$ means that an element is the successor of another element in the total order. Notice that both refer
to the next element of some order but on a different relation. In the sequel of the paper when me mean the first we
will use the term ``successor in $R$'' while for the second we will use the term ``successor build-in predicate''. When
we do not specify it should be evident from the context.
\\
\\
\noindent \textbf{Bottom-up evaluation and complexity }
\\
\\
\noindent The \textit{bottom-up evaluation} of a query, used in the proofs of the main theorems of this work,
initializes the IDB predicates to be empty and repeatedly applies the rules to add tuples to the IDB predicates, until
no new tuples can be added \cite{Ull88,AHV95,ZCF+97}. In stratified Datalog strata are used in order to structure the
computation in a bottom-up fashion. That is, the head predicates of a given stratum are evaluated only after all head
predicates of the lower strata have been computed. This way any negated subgoal is treated as if it were an EDB
relation.

    There are two main complexity measures for Datalog and its extensions.
\begin{itemize}
    \item   \textit{data complexity} which assumes a \textit{fixed} Datalog program and \textit{variable} input databases, and
    \item   \textit{program complexity} which refers to \textit{variable} Datalog programs over a \textit{fixed} input database.
\end{itemize}

In general, Datalog is P--complete with respect to data complexity and EXPTIME--complete with respect to program complexity \cite{Var82,Imm86}. Although there are different semantics for negation in Logic Programming (e.g., stratified negation, well-founded semantics, stable model semantics, etc.), for stratified programs these semantics coincide. Recall that a program is stratified if there is no recursion through negation. Stratified programs have a unique stable model which coincides with the stratified model, obtained by partitioning the program into an ordered number of strata and computing the fixpoints of every stratum in their order. Datalog with stratified negation is P--complete with respect to data complexity and EXPTIME--complete with respect to program complexity \cite{ABW88}. An excellent survey regarding these issues is \cite{DEG+01}.

\section{Embedding CTL to stratified Datalog}    \label{From CTL to STD}

In the present and next section  we establish that there is a fragment of stratified Datalog which has the same
expressive power as CTL. This fragment, which we define in  Subsection \ref{STD}, is called STD (for Stratified
Temporal Datalog). The following theorem is the result of the two main theorems of Sections 4 and 5 (Theorems
\ref{Thrm: From CTL to STD Soundness & Completeness} and \ref{Thrm: From STD to CTL Soundness & Completeness}) and it
states that CTL and STD have the same expressive power.

\begin{Theorem}   \label{expressivepower}
Consider the languages CTL and STD. The following hold.
\begin{enumerate}
 \item   Let $\mathcal{K}$ be a finite Kripke structure and $\varphi$  a CTL formula. Then there is a  relational database $D$ and a STD program $\Pi$
    such that the following holds:
    \begin{equation}
    \varphi [\mathcal{K}] = G_\Pi (D)   \label{Eq: From CTL to STD Soundness & Completeness1}
\end{equation}
Moreover  $D$ and  $\Pi$ are computed in time linear in the size of  $\mathcal{K}$ and $\varphi$.
 \item  Let $D$ be a  relational database  and  $\Pi$ a STD program. Then there is a finite Kripke structure $\mathcal{K}$ and a CTL formula  $\varphi$
    such that the following holds:
    \begin{equation}
     G_\Pi (D)=  \varphi [\mathcal{K}]  \label{Eq: From CTL to STD Soundness & Completeness2}
\end{equation}
Moreover  $\mathcal{K}$ and $\varphi$ are computed in time linear in the size of  $D$ and  $\Pi$.
\end{enumerate}
\end{Theorem}

\noindent We start by giving  the definition of the class STD in the following subsection together with some
properties.

\subsection{The class STD}                  \label{STD}

\subsubsection{Definition}

The programs of this class are built-up from: (a) a single binary predicate $R$ and an arbitrary number of unary EDB
predicates $P_0, \ldots, P_n$ , and (b) binary and unary IDB predicates. One unary IDB predicate is chosen to be the
goal predicate of the program.

\noindent The programs
            {\footnotesize  $G(x)  \longleftarrow P_i(x)$}
            and
            {\footnotesize  $\left\{\begin{array}{l}
                            G(x)  \longleftarrow W(x) \\
                            \Pi^n \\
                            \end{array}\right.$},
where {\footnotesize$\Pi^n$} is an abbreviation for
            {\footnotesize  $\left\{\begin{array}{l}
                            W(x)  \longleftarrow R(x,y)                                             \\
                            W(x)  \longleftarrow R(y,x)                                             \\
                            W(x)  \longleftarrow P_0(x)                                             \\
                            \dots                                                                   \\
                            W(x)  \longleftarrow P_n(x)                                             \\
                            \end{array}\right.$},
are STD programs having $G$ as the goal predicate. Inductively if $\Pi_1,\Pi_2$ are STD programs with goal predicates
$G_1$, $G_2$ respectively and with disjoint sets of IDB predicates (with the exception of $A$ and $W$ which are the
same in all programs) then  $\Pi$ is the union of the rules of $\Pi_1,\Pi_2$ and one of the following five sets of
rules -- predicate names $G$ and $B$ are new.

\begin{center}
{\footnotesize
\begin{tabular}{ll}
$\left\{\begin{array}{l}
                            G(x) \longleftarrow W(x), \neg G_1(x) \\
                            \Pi^n \\
\end{array}\right.$
& $\left\{\begin{array}{l}
                            G(x) \longleftarrow \neg A(x), G_1(x) \\
                            G(x) \longleftarrow R(x,y), G_1(y) \\
                            A(x) \longleftarrow R(x,y) \\
\end{array}\right.$
\\
$\left\{\begin{array}{l}
                            G(x) \longleftarrow G_1(x), G_2(x) \\
\end{array}\right.$
&
\\
$\left\{\begin{array}{l}
                            G(x) \longleftarrow G_2(x) \\
                            G(x) \longleftarrow G_1(x), R(x,y), G(y) \\
\end{array}\right.$
& $\left\{\begin{array}{l}
                            G(x) \longleftarrow G_1(x), G_{2}(x)                        \\
                            G(x) \longleftarrow G_2(x), \neg A(x)   \\
                            G(x) \longleftarrow B(x,x)   \\
                            G(x) \longleftarrow G_2(x), R(x,y), G(y)   \\
                            B(x,y) \longleftarrow G_2(x), R(x,y), G_2(y)                        \\
                            B(x,y) \longleftarrow G_2(x), R(x,u), B(u,y)                     \\
                            A(x) \longleftarrow R(x,y) \\
                            \end{array}\right. $
\\
\end{tabular}
}
\end{center}

Only the programs produced by the rules above are STD programs.

\subsubsection{Properties}

In the following paragraphs we provide some intuition about the IDB predicates of STD programs and we give a succinct
way to refer to STD programs which reflects their connection to CTL. Finally we show that STD programs are stratified.

Predicates $A$ and $W$ are auxiliary predicates denoting the ``ancestor'' relation and the ``domain'' respectively. The
intuition behind the IDB predicates $W, A$ and $B$, is the following:\\
$\bullet$   $W(x)$ as defined by $\Pi^n$ says that $x$ belongs to the domain of the database, i.e., appears in the relations that comprise the database. \\
$\bullet$   $A(x)$ asserts that state $x$ has at least one successor.\\
$\bullet$   $B(x, y)$ captures the notion of a path from state $x$ to state $y$, such that $G_2$ holds at every state along this path. In view of the fact that $G_2$ corresponds to a CTL formula (let's say $\psi_2$), $B(x, x)$ asserts the existence of a cycle having the property that $\psi_2$ holds at every state of this cycle.

For a more succinct presentation and for ease of reference we use the \textit{program operators} $\overline{[\cdot]}$,
$\bigwedge[\cdot, \cdot]$, $\mathbf{X}[\cdot]$, $\bigcup[\cdot, \cdot]$ and $\widetilde{\bigcup} [\cdot, \cdot]$
depicted in Figure~\ref{Fig: STD Query Operators}, where programs $\Pi_1$ and $\Pi_2$ are over disjoint sets of IDB
predicates (except $A$ and $W$ which are the same always) and $G$ and $B$ are new predicate names. It is useful to note
that using these operators, the class STD can be equivalently defined as follows:

\begin{Definition}      \label{Def: STD} \
\begin{itemize}
    \item   The programs
            {\footnotesize                              $G(x)  \longleftarrow P_i(x)$}
            and
            {\footnotesize  $\left\{\begin{array}{l}
                            G(x)  \longleftarrow W(x) \\
                            \Pi^n \\
                            \end{array}\right.$}
 are $\mathrm{STD}_n$ programs having $G$ as the goal predicate.
    \item   If $\Pi_1$ and $\Pi_2$ are $\mathrm{STD}_n$ programs with goal predicates $G_1$ and $G_2$ respectively, then $\overline{[\Pi_1]}$, $\bigwedge[\Pi_1, \Pi_2]$, $\mathbf{X}[\Pi_1]$, $\bigcup[\Pi_1, \Pi_2]$ and $\widetilde{\bigcup}[\Pi_1, \Pi_2]$ are also $\mathrm{STD}_n$ programs with goal predicate $G$.
    \item   The class $\mathrm{STD}$ is the union of the $\mathrm{STD}_n$ subclasses:
    \begin{equation}
        STD = \bigcup_{n \geq 0} STD_n
    \end{equation}
\end{itemize}
\end{Definition}

\begin{Example}
Consider the STD program $\Pi = \widetilde{\bigcup} [ \widetilde{\bigcup} [ \Pi_1, \Pi_2 ], \overline{[\Pi_3]}]$, where
$\Pi_1, \Pi_2$ and $\Pi_3$ are the simple STD programs $G_1(x) \longleftarrow P(x)$, $G_2(x) \longleftarrow Q(x)$ and
$G_3(x) \longleftarrow T(x)$, respectively. The rules comprising $\Pi$ are shown below $($$G_4$ and $G_5$ are the goals
of the subprograms $\widetilde{\bigcup} [ \Pi_1, \Pi_2 ]$ and $\overline{[\Pi_3]}$$):$

\begin{center}
{\footnotesize $\Pi =$  $\left\{\begin{array}{l}
                                        G(x)        \longleftarrow G_4(x), G_5(x)               \\
                                        G(x)        \longleftarrow G_5(x), \neg A(x)            \\
                                        G(x)        \longleftarrow B(x, x)                      \\
                                        G(x)        \longleftarrow G_5(x), R(x, y), G(x)        \\
                                        B(x, y)     \longleftarrow G_5(x), R(x, y), G_5(y)      \\
                                        B(x, y)     \longleftarrow G_5(x), R(x, u), B(u, y)     \\
                                        G_4(x)      \longleftarrow G_1(x), G_2(x)               \\
                                        G_4(x)      \longleftarrow G_2(x), \neg A(x)            \\
                                        G_4(x)      \longleftarrow B_1(x, x)                    \\
                                        G_4(x)      \longleftarrow G_2(x), R(x, y), G_4(x)      \\
                                        B_1(x, y)   \longleftarrow G_2(x), R(x, y), G_2(y)      \\
                                        B_1(x, y)   \longleftarrow G_2(x), R(x, u), B_1(u, y)   \\
                                        G_5(x)      \longleftarrow W(x), \neg G_3(x)            \\
                                        G_1(x)      \longleftarrow P(x)                         \\
                                        G_2(x)      \longleftarrow Q(x)                         \\
                                        G_3(x)      \longleftarrow T(x)                         \\
                                        A(x)        \longleftarrow R(x, y)                      \\
                                        \Pi^n
\end{array}\right. $
}
\end{center}
\hfill $\blacktriangle$
\end{Example}

\begin{figure}[!htb]
\centering
\textbf{The query operators of the class STD} \\
{\scriptsize
\begin{tabular} {l}
\\
\hspace{0.7 cm} $\Pi^n =                        \left\{\begin{array}{l}
                                                W(x)  \longleftarrow R(x,y)                                             \\
                                                W(x)  \longleftarrow R(y,x)                                             \\
                                                W(x)  \longleftarrow P_0(x)                                             \\
                                                \dots                                                                   \\
                                                W(x)  \longleftarrow P_n(x)                                             \\
                                                \end{array}\right.$
\\
\hspace{0.6 cm} $\overline{[\Pi_1]} =           \left\{\begin{array}{l}
                                                G(x) \longleftarrow W(x), \neg G_1(x)                                   \\
                                                \Pi_1                                                                   \\
                                                \Pi^n                                                                   \\
                                                \end{array}\right.$
\\
$\bigwedge[\Pi_1, \Pi_2] =                      \left\{\begin{array}{l}
                                                G(x) \longleftarrow G_1(x), G_2(x)                                      \\
                                                \Pi_1                                                                   \\
                                                \Pi_2                                                                   \\
                                                \end{array}\right.$
\\
\hspace{0.4 cm}  $\mathbf{X}[\Pi_1] =            \left\{\begin{array}{l}
                                                G(x) \longleftarrow G_1(x), \neg A(x)                                   \\
                                                G(x) \longleftarrow R(x,y), G_1(y)                                      \\
                                                A(x) \longleftarrow R(x,y)                                              \\
                                                \Pi_1                                                                   \\
                                                \end{array}\right.$
\\
$\bigcup[\Pi_1, \Pi_2] =                        \left\{\begin{array}{l}
                                                G(x) \longleftarrow G_2(x)                                              \\
                                                G(x) \longleftarrow G_1(x), R(x,y), G(y)                                \\
                                                \Pi_1                                                                   \\
                                                \Pi_2                                                                   \\
                                                \end{array}\right.$
\\
$\widetilde{\bigcup}[\Pi_1, \Pi_2] =            \left\{\begin{array}{l}
                                                G(x) \longleftarrow G_1(x), G_{2}(x)                                    \\
                                                G(x) \longleftarrow G_2(x), \neg A(x)                                   \\
                                                G(x) \longleftarrow B(x,x)                                              \\
                                                G(x) \longleftarrow G_2(x), R(x,y), G(y)                                \\
                                                B(x,y) \longleftarrow G_2(x), R(x,y), G_2(y)                            \\
                                                B(x,y) \longleftarrow G_2(x), R(x,u), B(u,y)                            \\
                                                A(x) \longleftarrow R(x,y)                                              \\
                                                \Pi_1                                                                   \\
                                                \Pi_2                                                                   \\
                                                \end{array}\right. $
\end{tabular}
} \caption{\textit{\small{These are the query operators used in the definition of the class STD. $\Pi_1$ and $\Pi_2$
are $\mathrm{STD}_n$ programs with goal predicates $G_1$ and $G_2$ respectively. $G$ and $B$ are ``fresh" predicate
symbols, i.e., they do not appear in $\Pi_1$ or $\Pi_2$. In contrast, $A$ and $W$ are the same in all programs. $\Pi^n$
is a convenient abbreviation of the rules depicted here.}}} \label{Fig: STD Query Operators}
\end{figure}

The following proposition proves that the STD class is a fragment of stratified Datalog.

\begin{Proposition}
    Every STD program is stratified.
\end{Proposition}
\textbf{Proof}\\
Given that $\Pi_1$, $\Pi_2$ are stratified programs, any set of rules that might be added to $\Pi_1$, $\Pi_2$ in order to form program $\Pi$ according to Definition \ref{Def: STD} preserves the stratification of the program. \hfill$\dashv$

\subsection{From CTL formulae to relational queries}

Embedding CTL into STD amounts to defining a mapping $\mathbf{h} = (h_f, h_s)$ such that:

\begin{enumerate}
    \item   $h_f$ maps CTL formulae into STD programs, that is given a formula $\varphi$, $h_f(\varphi)$ is a program $\Pi$ with unary goal predicate $G$.
    \item   $h_s$ maps temporal Kripke structures (on which CTL formulae are interpreted) to relational databases, i.e., $h_s(\mathcal{K})$ is a database $D$.
    \item   For this mapping it holds:
\begin{displaymath}
    \varphi [\mathcal{K}] = G_\Pi (D), \textrm{where} \ \Pi = h_f(\varphi) \ \textrm{and} \ D = h_s(\mathcal{K})
\end{displaymath}
\end{enumerate}

The correspondence of CTL formulae to Datalog programs is the core of both translations. The exact mapping $h_f$ of CTL
formulae into STD programs is given below. Note that we use the operators of Figure \ref{Fig: STD Query Operators} to
facilitate the reading and that $\Pi_i$ corresponds to subformula $\psi_i, i = 1, 2$.

\begin{Definition}      \label{Def: CTL to STD Translation}
    Let $\varphi$ be a CTL formula and let $p_0, \ldots, p_n$ be the atomic propositions appearing in $\varphi$. Then $h_f (\varphi)$ is the $\mathrm{STD}_n$ program defined recursively as follows:
      \begin{enumerate}
        \item   If $\varphi \equiv p_i$ or $\varphi \equiv \top$, then $h_f (\varphi)$ is
                {\footnotesize  $\left\{\begin{array}{l}
                                G(x)  \longleftarrow P_i(x) \\
                                \end{array}\right.$}
                and
                {\footnotesize  $\left\{\begin{array}{l}
                                G(x)  \longleftarrow W(x) \\
                                \Pi^n \\
                                \end{array}\right.$}
                , respectively.
        \item   If \ $\varphi \equiv \neg \psi_1$ or $\varphi \equiv \psi_1 \wedge \psi_2$, then $h_f (\varphi)$ is $\overline{[\Pi_1]}$ and $\bigwedge[\Pi_1, \Pi_2]$, respectively.
        \item   If $\varphi \equiv \textbf{E} \nextt \psi_1$ or $\varphi \equiv \textbf{E}(\psi_1 \mathbf{U} \psi_2)$ or $\varphi \equiv \textbf{E}(\psi_1 \widetilde{\mathbf{U}} \psi_2)$, then $h_f (\varphi)$ is $\mathbf{X}[\Pi_1]$, $\bigcup[\Pi_1, \Pi_2]$ and $\widetilde{\bigcup}[\Pi_1, \Pi_2]$, respectively. \hfill \blackrec
\end{enumerate}
\end{Definition}

\noindent The following example illustrates the translation presented above.

\begin{Example}
   Let us consider a CTL formula $\varphi$ that contains the modality $\widetilde{\mathbf{U}}$, e.g., $\neg \mathbf{E} (\psi_1 \widetilde{\mathbf{U}} \psi_2)$. Then
\begin{center}
{\footnotesize
 $h_f (\varphi) =$  $\left\{\begin{array}{l}
                                        G(x)    \longleftarrow W(x), \neg G_1(x)        \\
                                        G_1(x)  \longleftarrow G_2(x), G_3(x)           \\
                                        G_1(x)  \longleftarrow G_3(x), \neg A(x)        \\
                                        G_1(x)  \longleftarrow B(x,x)                   \\
                                        G_1(x)  \longleftarrow G_3(x), R(x,y), G_1(y)   \\
                                        B(x,y)  \longleftarrow G_3(x), R(x,y), G_3(y)   \\
                                        B(x,y)  \longleftarrow G_3(x), R(x,u), B(u,y)   \\
                                        A(x)    \longleftarrow R(x,y)                   \\
                                        \Pi_2                                           \\
                                        \Pi_3                                           \\
                                        \Pi^n
                                        \end{array}\right. $

}
\end{center}
\noindent   where $G_2$, $G_3$, $\Pi_2$ and $\Pi_3$ are the goal predicates and the programs that correspond to subformulae $\psi_1$ and $\psi_2$, respectively.    \hfill $\blacktriangle$
\end{Example}

    The construction of STD programs that correspond to CTL formulae can be performed efficiently. This is formalized by the next proposition; its proof is a direct consequence of Definition \ref{Def: CTL to STD Translation}.

\begin{Proposition}  \label{Prop: Linear Construction of STD Queries}  \
        Given a CTL formula $\varphi$, the corresponding STD program $\Pi$, which is of size $O(|\varphi|)$, can be constructed in time $O(|\varphi|)$.
\end{Proposition}

\subsection{From finite Kripke structures to relational databases}

In this section we show how finite Kripke structures can be seen as relational databases. Definition \ref{Def: From Kripke Structures to DBs} states formally the details of this mapping.

\begin{Definition}      \label{Def: From Kripke Structures to DBs}
    Let $AP$ be a finite set $\{p_0, \ldots, p_n\}$ of atomic propositions and assume that $\mathcal{K} = \ang{W, R, V}$ is a finite Kripke structure for $AP$. Then $h_s (\mathcal{K})$ is the database $\ang{R, P_0, \ldots, P_n}$, where $P_i = \{ s \in W \mid p_i \in V(s) \}$ contains the states at which $p_i$ is true $(0 \leq i \leq n)$.

    Further, to $\mathcal{K}$ corresponds the database schema $\mathfrak{D}_\mathcal{K} = \ang{W, R, P_0, \ldots, P_n}$, with domain the set of states $W$, one binary predicate symbol $R$ and an arbitrary number of unary predicate symbols $P_0, \ldots, P_n$.\footnote{As we have already pointed out, for simplicity we use the same notation, e.g., $R, P_0, \ldots, P_n$ both for the predicate symbols and the relations. The context makes clear whether $R, P_0, \ldots, P_n$ stand for predicate symbols or relations.} A database schema of this form, i.e., containing a \emph{single binary} predicate symbol and having all other \emph{unary}, is called a \emph{Kripke schema}. \hfill \blackrec
\end{Definition}

\noindent    The following proposition is a straightforward consequence of Definition \ref{Def: From Kripke Structures to DBs}.

\begin{Proposition}     \label{Prop: From Kripke Structures to DBs}
A finite Kripke structure $\mathcal{K}$ can be converted into a relational database $D = h_s (\mathcal{K})$ of size
$O(|\mathcal{K}|)$\footnote{The number $n$ of the unary relations $P_0, \ldots, P_n$ is a constant of the problem.} in
time $O(|\mathcal{K}|)$.
\end{Proposition}

    Notice that the relation $R$ of $h_s (\mathcal{K})$ is \emph{total}. Moreover, every path $s_0, s_1, s_2, \ldots$ of $\mathcal{K}$ gives rise to the path $s_0, s_1, s_2, \ldots \ $ in $h_s (\mathcal{K})$ and vice versa: if $s_0, s_1, s_2, \ldots$ is a path in $h_s (\mathcal{K})$, then
\begin{equation}
    R(s_i, s_{i+1}), \mathrm{ \ for \ every \ } \ i\geq 0   \label{Eq: Path Connectivity (STD)}
\end{equation}

\noindent    The next proposition states formally the basic property of $B(x, x)$.

\begin{Proposition}     \label{Prop: Existential TC} \
    $B(s, s)$ holds iff there exists a finite sequence of states $s_0, \ldots, s_n$ in $D_\mathcal{K}$ such that $s_0 = s_n = s$ and
$G_2(s_i)$, for every $i, \ 0 \leq i \leq n$.
\end{Proposition}

    In the proof of the main result in this section (Theorem \ref{Thrm: From CTL to STD Soundness & Completeness}) we need the next proposition, which is basically just a simple application of the pigeonhole principle.

\begin{Proposition}     \label{Prop: Pigeonhole Principle in K}
    Let $\mathcal{K} = \ang{W, R, V}$ be a finite Kripke structure and let $s_0$, \ldots, $s_i$, \ldots, $s_j$, \ldots,
$s_n$ be a finite path in $\mathcal{K}$, where $n \geq |W|$. Then, there exists a state $s \in W$ such that $s_i = s_j
= s$.
\end{Proposition}

\subsection {Embedding CTL into STD}

We are ready now to prove the main result of this section, which asserts that the mapping from CTL formulae to STD
programs we defined earlier.

\begin{Theorem}         \label{Thrm: From CTL to STD Soundness & Completeness}
    Let $\mathcal{K}$ be a finite Kripke structure and let $D$ be the corresponding relational database. If $\varphi$ is a
CTL formula and $\Pi$ its corresponding STD program $($see Definition \ref{Def: CTL to STD Translation}$)$, then the
following holds:
\begin{equation}
    \varphi [\mathcal{K}] = G_\Pi (D)   \label{Eq: From CTL to STD Soundness & Completeness}
\end{equation}
\end{Theorem}

\noindent\textbf{Proof}\\
To facilitate the readability of this proof, we use subscripts in the goal predicates to denote the corresponding CTL
subformulae. That is we write $G_{\mathbf{E} \nextt \psi}$ to denote that $G$ is the goal predicate of the program
corresponding to $\mathbf{E} \nextt \psi$. We prove that (\ref{Eq: From CTL to STD Soundness & Completeness}) holds by
simultaneous induction on the structure of formula $\varphi$.

\begin{enumerate}
    \item   If $\varphi\equiv p$, where $p\in AP$, or $\varphi \equiv \top$, then the corresponding programs are those of
Definition \ref{Def: CTL to STD Translation}.(1):
        \begin{itemize}
            \item               ${\mathcal K}, s \models p \Leftrightarrow p \in V(s) \Leftrightarrow P(s)$ is a ground fact of $D \Leftrightarrow s \in G_p(D)$.
            \item               $(\Rightarrow)$ ${\mathcal K}, s \models \top \Rightarrow s \in W \Rightarrow$ (by the totality of $R$) there exists $t \in W$ such that $(s, t) \in R$ $\Rightarrow s \in W_{\Pi^n}(D) \Rightarrow s \in G_\top(D)$. \\
                                $(\Leftarrow)$  $s \in G_\top(D) \Rightarrow s \in W_{\Pi^n}(D) \Rightarrow s$ appears in one of $R, P_0, \ldots, P_n \Rightarrow s \in W \Rightarrow {\mathcal K}, s \models \top$.
        \end{itemize}
    \item   If $\varphi\equiv \neg \psi$ or $\varphi\equiv\psi_1\wedge\psi_2$, then the corresponding programs are shown in Definition \ref{Def: CTL to STD Translation}.(2).
        \begin{itemize}
            \item[$\neg:$]      $(\Rightarrow)$ ${\mathcal K}, s \models \varphi \Rightarrow {\mathcal K}, s \models \neg \psi \Rightarrow {\mathcal K}, s \not \models \psi \Rightarrow$ (by the induction hypothesis) $s \not \in G_{\psi} (D) \Rightarrow s \in G_{\varphi} (D)$. \\
                                $(\Leftarrow)$  $s \in G_{\varphi} (D) \Rightarrow s \not \in G_{\psi} (D) \Rightarrow$ (by the induction hypothesis) ${\mathcal K}, s \not \models \psi \Rightarrow {\mathcal K}, s \models \neg \psi \Rightarrow {\mathcal K}, s \models \varphi$.

            \item[$\wedge:$]    $(\Rightarrow)$ ${\mathcal K}, s \models \varphi \Rightarrow {\mathcal K}, s \models \psi_1$ and ${\mathcal K}, s \models \psi_2  \Rightarrow$ (by the induction hypothesis) $s \in G_{\psi_1} (D)$ and $s \in G_{\psi_2} (D) \Rightarrow s \in G_{\psi_1}(D) \cap G_{\psi_2}(D) \Rightarrow s\in G_{\varphi} (D)$. \\
                                $(\Leftarrow)$  $s \in G_{\varphi} (D) \Rightarrow s\in G_{\psi_1}(D) \cap G_{\psi_2}(D) \Rightarrow s \in G_{\psi_1} (D)$ and $s \in G_{\psi_2} (D) \Rightarrow$ (by the induction hypothesis) ${\mathcal K}, s \models\psi_1$ and ${\mathcal K}, s \models\psi_2 \Rightarrow {\mathcal K}, s \models \varphi$.
        \end{itemize}
    \item   If $\varphi \equiv \mathbf{E} \nextt \psi$, then the corresponding program is that of Definition \ref{Def: CTL to STD Translation}.(3).
        \begin{itemize}
            \item[$(\Rightarrow)$]  ${\mathcal K}, s \models \mathbf{E} \nextt \psi \Rightarrow$ there exists a path $\pi = s_0, s_1, s_2, \ldots$ \ with initial state $s_0 = s$, such that $\mathcal{K}, \pi \models \nextt \psi \Rightarrow$ ${\mathcal K}, \pi^1 \models \psi$ for the path $\pi^1 = s_1, s_2, \ldots$ $\Rightarrow$ ${\mathcal K}, s_1 \models \psi$ $\Rightarrow$ (by the induction hypothesis) $s_1 \in G_{\psi} (D)$. Furthermore, from (\ref{Eq: Path Connectivity (STD)}) we know that $R(s_0, s_1)$ holds. From the second rule of $\Pi_\varphi$, by combining $G_{\psi}(s_1)$ with $R(s_0, s_1)$, we derive $G_\varphi (s_{0})$ and, thus, $s_{0} \in G_\varphi (D)$.

            \item[$(\Leftarrow)$]   Let us assume that $s \in G_\varphi (D)$. From the rules of $\Pi_\varphi$\footnote{Recall that in this case relation $R$ is total. Hence, $s \in A(D)$ and the first rule does not add new states to $G_\varphi (D)$.}
there exists a $s_1$ such that $R(s, s_1)$ and $G_{\psi} (s_1)$ hold. By the induction hypothesis we get ${\mathcal K},
s_1 \models \psi$. Let $\pi = s_0, s_1, s_2, \ldots$ be any path with initial state $s_0 = s$ and second state $s_1$.
Clearly, then ${\mathcal K}, \pi^1 \models \psi \Rightarrow {\mathcal K}, \pi \models \nextt \psi \Rightarrow {\mathcal
K}, s \models \varphi$.
        \end{itemize}
    \item   If $\varphi \equiv \mathbf{E} (\psi_1 \mathbf{U} \psi_2)$, then the corresponding program is that of Definition
\ref{Def: CTL to STD Translation}.(3).
        \begin{itemize}
            \item[$(\Rightarrow)$]  ${\mathcal K}, s \models \mathbf{E} (\psi_1 \mathbf{U} \psi_2) \Rightarrow$ there exists a path $\pi = s_0, s_1, s_2,
\ldots$ \ with initial state $s_0 = s$, such that ${\mathcal K}, \pi^i \models \psi_2$ and ${\mathcal K}, \pi^j \models
\psi_1$ $(0 \leq j \leq i-1)$ $\Rightarrow$ ${\mathcal K}, s_i \models \psi_2$ and ${\mathcal K}, s_j \models \psi_1$
$(0 \leq j \leq i-1)$ $\Rightarrow$ $s_i \in G_{\psi_{2}} (D)$ and $s_j \in G_{\psi_{1}} (D)$ $(0\leq j\leq i-1)$ (by
the induction hypothesis). From  (\ref{Eq: Path Connectivity (STD)}) we know that $R(s_r, s_{r+1})$, $0 \leq r < i$.
From the first rule of $\Pi_{\varphi}: G_\varphi (x) \longleftarrow G_{\psi_2} (x)$ we derive that $G_{\varphi} (s_i)$.
Successive applications of the second rule of $\Pi_{\varphi}: G_\varphi (x) \longleftarrow G_{\psi_1} (x)$, $R (x, y)$,
$G_\varphi (y)$ yield $G_\varphi (s_{i-1})$, $G_\varphi (s_{i-2})$, \ldots, $G_\varphi (s_1)$, $G_\varphi (s_0)$. Thus,
$s_0 \in G_\varphi (D)$.

            \item[$(\Leftarrow)$]   For the inverse direction, suppose that $s \in G_\varphi (D)$. The rules of $\Pi_\varphi$ imply the existence of a
state $s_i$ (possibly $s_i = s$) such that $G_{\psi_2} (s_i)$. In addition, there exists a sequence of states $s_0 = s,
s_1, \ldots, s_i$ such that $R(s_r, s_{r+1})$ and $G_{\psi_1} (s_r)$ ($0 \leq r < i$). By the induction hypothesis we
get that ${\mathcal K}, s_i \models \psi_2$ and ${\mathcal K}, s_j \models \psi_1$ $(0 \leq j \leq i-1)$ (because
$\psi_1$ and $\psi_2$ are state formulae). Let $\pi = s_0, s_1, s_2, \ldots, s_i, \ldots$ be any path with initial
segment $s_0, s_1, \ldots, s_i$. Then, ${\mathcal K}, \pi^i \models \psi_2$ and ${\mathcal K}, \pi^j \models \psi_1$
$(0 \leq j \leq i-1)$, i.e., ${\mathcal K}, \pi \models \varphi$.
        \end{itemize}
    \item   If $\varphi \equiv \mathbf{E} (\psi_1 \mathbf{\widetilde{U}} \psi_2)$, then the corresponding program is that of Definition \ref{Def: CTL to STD Translation}.(3).
        \begin{itemize}
            \item[$(\Rightarrow)$]  Recall from Section 2 that ${\mathcal K}, s \models \mathbf{E} (\psi_1 \mathbf{\widetilde{U}} \psi_2)$ means that there exists a path $\pi = s_0, s_1, s_2, \ldots$ \ with initial state $s_0 = s$, such that either (1) ${\mathcal K}, \pi^i \models \psi_2$, for every $i \geq 0$, or (2) ${\mathcal K}, \pi^i \models \psi_1 \wedge \psi_2$ and ${\mathcal K}, \pi^j \models \psi_2, \ 0 \leq j \leq i-1$. We examine both cases:

                (a)     In the first case ${\mathcal K}, s_i \models \psi_2$, for every $i \geq 0$. The induction hypothesis gives that $s_i \in G_{\psi_2} (D)$, for every $i \geq 0$. Let $s_0, s_1, s_2, \ldots, s_n$ be an initial segment of $\pi$, with $n \geq |W|$. From Proposition \ref{Prop: Pigeonhole Principle in K} we know that in the aforementioned sequence there exists a state $t$ such that $t = s_k = s_l$, $0 \leq k < l \leq n$. Then Proposition \ref{Prop: Existential TC} implies that $(s_k, s_k) \in B(D)$. From the third rule of $\Pi_\varphi$:  $G_\varphi(x) \longleftarrow B(x, x)$, we derive that $G_\varphi(s_k)$. Successive applications of the fourth rule of $\Pi_\varphi$: $G_\varphi(x) \longleftarrow G_{\psi_2} (x), R(x, y), G_\varphi(y)$ yield $G_\varphi(s_{k-1})$, $G_\varphi(s_{k-2})$, \ldots, $G_\varphi(s_1)$, $G_\varphi(s_0)$. Accordingly, $s_0 \in G_\varphi(D)$.

                (b)     In the second case, ${\mathcal K}, s_i \models \psi_1 \wedge \psi_2$ and ${\mathcal K}, s_j \models
\psi_2$, $0 \leq j \leq i-1$. By the induction hypothesis $s_i \in G_{\psi_1} (D)$ and $s_j \in G_{\psi_2} (D)$, $0
\leq j \leq i$. From the first rule of $\Pi_\varphi$: $G_\varphi(x) \longleftarrow G_{\psi_1} (x), G_{\psi_2} (x)$, we
derive that $G_\varphi(s_i)$. Successive applications of the fourth rule of $\Pi_\varphi$: $G_\varphi(x) \longleftarrow
G_{\psi_2} (x), R(x, y), G_\varphi(y)$ yield $G_\varphi(s_{i-1})$, $G_\varphi(s_{i-2})$, \ldots, $G_\varphi(s_1)$,
$G_\varphi(s_0)$. Therefore, $s_0 \in G_\varphi(D)$.

            \item[$(\Leftarrow)$]   For the inverse direction, suppose that $s_0 \in G_\varphi(D)$. We define $G_\varphi (D, n)$ to be the set of ground
facts of the IDB $G_\varphi$ that have been computed during the first $n$ rounds of the evaluation of the last stratum
of the program $\Pi_\varphi$. We shall prove that for every $t \in G_\varphi (D, n)$, there exists a path $\pi = t_0,
t_1, t_2, \ldots$ \ with initial state $t_0 = t$, such that ${\mathcal K}, \pi \models \varphi$. We use induction on
the number of rounds $n$.

    (a) \hspace{0.2 cm}     If $n = 1$, then $t$ must appear either due to the first rule of $\Pi_\varphi$: $G_\varphi(x) \longleftarrow G_{\psi_1}
(x), G_{\psi_2} (x)$ or due to the third rule of $\Pi_\varphi$: $G_\varphi(x) \longleftarrow B(x, x)$\footnote{Relation
$R$ is total, meaning that $t \in A_\varphi(D)$, and, thus, $t$ could not have appeared from an application of the
second rule.}, assuming $B$ is in a previous stratum. Note that if $B$ is in the last stratum, then, of course, $t$
could not have appeared due to the third rule. In the former case $t \in G_{\psi_1} (D_\mathcal{K}) \cap G_{\psi_2}
(D_\mathcal{K})$; the induction hypothesis for $\psi_1$ and $\psi_2$ means that ${\mathcal K}, t \models \psi_1 \wedge
\psi_2$, which immediately implies that ${\mathcal K}, \pi \models \varphi$ for any path $\pi = t_0, t_1, t_2, \ldots$
with initial state $t_0 = t$. In the latter case, $(t, t) \in B_\varphi(D)$ and, in view of Proposition \ref{Prop:
Existential TC}, this implies the existence of a finite sequence $t_0, t_1, \ldots, t_k$, such that $t_0 = t_k = t$ and
${\mathcal K}, t_j \models \psi_2$, \ $0 \leq j \leq k$. Consider the path $\pi = (t_0, t_1, \ldots, t_k)^\omega$; for
this path ${\mathcal K}, \pi \models \varphi$.

    (b) \hspace{0.2 cm}     We show now that the claim holds for $n+1$, assuming that it holds for $n$. Suppose that $t$ first appeared in $G_\varphi (D, n+1)$ during round $n+1$. This could have happened either because of the third rule:  $G_\varphi(x) \longleftarrow B(x, x)$ or because of the fourth rule: $G_\varphi(x) \longleftarrow G_{\psi_2} (x), R(x, y), G_\varphi(y)$.

    In the first case $(t, t) \in B_\varphi(D)$. Then Proposition \ref{Prop: Existential TC} asserts the existence of a
finite sequence $t_0, t_1, \ldots, t_k$ of states, such that $t_0 = t_k = t$ and ${\mathcal K}, t_j \models \psi_2$, \
$0 \leq j \leq k$. Consider the path $\pi = (t_0, t_1, \ldots, t_k)^\omega$; for this path ${\mathcal K}, \pi \models
\varphi$.

    In the second case, we know that $G_{\psi_2}(t)$ and that there exists a $t_1$ such that $R(t, t_1)$ and
$G_\varphi(t_1)$. By the induction hypothesis, we get that ${\mathcal K}, t \models \psi_2$ and that ${\mathcal K}, t_1
\models \varphi$. Immediately then we conclude that ${\mathcal K}, \pi \models \varphi$, for the path $\pi = t_0, t_1,
t_2, \ldots$ \ with $t_0 = t$. \hfill$\dashv$
        \end{itemize}
\end{enumerate}

\section{Embedding a fragment of Stratified Datalog into CTL}       \label{From STD to CTL}

In the previous section we defined a mapping from CTL to the class of STD programs. In this section we work on the opposite direction, that is we define an embedding from STD to CTL. We start with explaining the technical challenges of this embedding.

\subsection{Technical Challenges}

In Kripke structures the accessibility relation $R$ is total and as a result the corresponding relational database contains a total binary relation $R$. Here lies the main problem when going from databases to Kripke structures: a database relation is not necessarily total. To overcome this problem we define the \emph{total closure} $R^t$ of an arbitrary binary relation $R$ with respect to a domain $W$ as follows:

\begin{equation}
    R^t = R \cup \{(x, x) \ | \ x \in W \mathrm{\ and} \not \exists y \mathrm{\ such \ that \ } R(x, y) \}
     \label{Eq: Total Closure (STD)}
\end{equation}

    In simple words the above equation means that even when $R$ is not total, we can still get a total relation by adding a self loop to the states that have no successors. Note that if $R$ is already total then $R^t = R$.

\subsection{From STD programs to CTL formulae}

We define a mapping $\mathbf{f} = (f_q, f_d)$ such that:

\begin{enumerate}
    \item   $f_q$ maps STD programs into CTL formulae, that is given a program $\Pi$ with unary goal predicate $G$, $f_q(\Pi)$ is a CTL formula $\varphi$.
    \item   $f_d$ maps relational databases to Kripke structures, i.e., $f_d(D)$ is a Kripke structure $\mathcal{K}$.
    \item   For this mapping it holds:
\begin{displaymath}
    G_\Pi (D) = \varphi [\mathcal{K}], \textrm{where} \ \varphi = f_q(\Pi) \ \textrm{and} \ \mathcal{K} = f_d(D)
\end{displaymath}
\end{enumerate}

    The correspondence of STD programs to CTL formulae is given below (subformula $\psi_i$ corresponds to subprogram $\Pi_i, i = 1, 2$).

\begin{Definition}      \label{Def: STD to CTL Translation}
    Given a $\mathrm{STD}_n$ program $\Pi$, $f_q (\Pi)$ is the CTL formula defined recursively as follows:
      \begin{enumerate}
        \item   If $\Pi = $ {\footnotesize  $\left\{\begin{array}{l}
                                            G(x)  \longleftarrow P_i(x) \\
                                            \end{array}\right.$}
                or $\Pi = $ {\footnotesize  $\left\{\begin{array}{l}
                                            G(x)  \longleftarrow W(x) \\
                                            \Pi^n \\
                                            \end{array}\right.$},
                then $f_q (\Pi)$ is $p_i$ and $\top$, respectively.
        \item   If \ $\Pi = \overline{[\Pi_1]}$ or $\Pi = \bigwedge[\Pi_1, \Pi_2]$, then $f_q (\Pi)$ is $\neg \psi_1$ and $\psi_1 \wedge \psi_2$, respectively.
        \item   If \ $\Pi = \mathbf{X}[\Pi_1]$ or $\Pi = \bigcup[\Pi_1, \Pi_2]$ or $\Pi = \widetilde{\bigcup}[\Pi_1, \Pi_2]$, then $f_q (\Pi)$ is $\textbf{E} \nextt \psi_1$, $\textbf{E}(\psi_1 \mathbf{U} \psi_2)$ and $\textbf{E}(\psi_1 \widetilde{\mathbf{U}} \psi_2)$, respectively. \hfill \blackrec
\end{enumerate}
\end{Definition}

    The following proposition asserts that the construction of CTL formulae that correspond to STD programs can be performed efficiently. Its proof is an immediate consequence of Definition \ref{Def: STD to CTL Translation}.

\begin{Proposition}  \label{Prop: Linear Construction of CTL Formulae}  \
        Given a STD program $\Pi$, the corresponding CTL formula $\varphi$, which is of size $O(|\Pi|)$, can be constructed in time $O(|\Pi|)$ .
\end{Proposition}

\subsection{From databases to finite Kripke structures}

In this section we show how an arbitrary relational database can be transformed into a finite Kripke structure in a
meaningful way. Definition \ref{Def: From DBs to Kripke Structures} has the details of this transformation.

\begin{Definition}  \label{Def: From DBs to Kripke Structures}
    Let $D$ be a database over the Kripke schema $\mathfrak{D}_\mathcal{K} = \ang{U, R, P_0, \ldots, P_n}$. We define the
\emph{domain} $W$ of $D$ as follows:

\begin{equation}
    W = \{ x \in U \ | \ R(x, y) \} \bigcup \{ x \in U \ | \ R(y, x) \} \bigcup_{i = 0}^n \{ x \in U \ | \ P_i(x) \}
    \label{Eq: The Domain of a Relational Database}
\end{equation}

    Let $D^t$ be the \emph{total} database $\ang{R^t, P_0, \ldots, P_n}$, where $R^t$ is the total closure of $R$ with
respect to $W$; then $f_d (D)$ is the finite Kripke structure $\ang{W, R^t, V}$ for $AP = \{p_0, \ldots, p_n\}$, with
$V(s) = \{ p_i \in AP \mid P_i(s) \}$. \hfill \blackrec
\end{Definition}

$f_d (D)$ is well-defined because $R^t$ is total as required by Definition~\ref{Def: Kripke Structure}. The next
proposition follows directly from Definition \ref{Def: From DBs to Kripke Structures}.

\begin{Proposition}     \label{Prop: From DBs to Kripke Structures}
    Let $D$ be a relational database $\ang{R, P_0, \ldots, P_n}$ over a Kripke schema and let $W$ be the domain of $D$ as
defined by $(\ref{Eq: The Domain of a Relational Database})$. $D$ can be transformed into a finite Kripke structure
$\mathcal{K} = f_d (D)$ of size $O(|W| + |R|) = O(|D|)$ in time $O(|D|)$.\footnote{Recall that the number $n$ of the
unary relations $P_0, \ldots, P_n$ is a constant of the problem.}
\end{Proposition}

    The main result of this section is that the mapping $\mathbf{f} = (f_q, f_d)$ is such the following holds: $G_\Pi (D) =
\varphi [\mathcal{K}]$, where $\varphi = f_q(\Pi)$ and $\mathcal{K} = f_d(D)$. Before proving that, we show that
$\mathrm{STD}$ programs can not distinguish between a database $D$ and the corresponding total database $D^t$, i.e.,
are invariant under total closure.

\begin{Theorem}         \label{Thrm: STD Query Equivalence under Total Closure}
    If $\Pi$ is a STD program with goal predicate $G$ and $D$ a database with a Kripke schema, then
\begin{equation}
    G_\Pi(D) = G_\Pi(D^t)       \label{Eq: STD Query Equivalence under Total Closure}
\end{equation}
\end{Theorem}

\noindent\textbf{Proof}\\
    We prove that (\ref{Eq: STD Query Equivalence under Total Closure}) holds by induction on the structure of the program $\Pi$.

\begin{enumerate}
        \item   If $\Pi = $ {\footnotesize  $\left\{\begin{array}{l}
                                            G(x)  \longleftarrow P_i(x) \\
                                            \end{array}\right.$},
                then $s \in G_\Pi(D) \Leftrightarrow P_i(s)$ is a ground fact of $D$ $\Leftrightarrow P_i(s)$ is a ground fact of $D^t$ $\Leftrightarrow s \in G_\Pi(D^t)$.
        \item   If $\Pi = $ {\footnotesize  $\left\{\begin{array}{l}
                                            G(x)  \longleftarrow W(x) \\
                                            \Pi^n \\
                                            \end{array}\right.$},
                then:
        \begin{itemize}
            \item[$(\Rightarrow)$]  $s \in W_{\Pi^n}(D) \Rightarrow$ $D$ contains a ground fact of the form $P_i(s)$ or $R(s, t)$ or $R(t, s)$. Obviously, $D^t$ also contains this ground fact, which means that $s \in W_{\Pi^n}(D^t)$.

            \item[$(\Leftarrow)$]   $s \in W_{\Pi^n}(D^t) \Rightarrow$ $D^t$ contains a ground fact of the form $P_i(s)$ or $R(s, t)$ or $R(t, s)$, or $R(s, s)$. In the first three cases $D$ also contains this ground fact; however, $D$ may not contain a ground fact of $D^t$ that has the form $R(s, s)$. If $D$ does not contain $R(s, s)$, this implies (recall the definition of $R^t$) that $D$ contains a fact $P_i(s)$ or a fact $R(t, s)$ for some constant $t$, but does not contain any fact of the form $R(s, u)$. But then we would have that $s \in W_{\Pi^n}(D)$ due to $P_i(s)$ or $R(t, s)$.
        \end{itemize}
        \item   If \ $\Pi = \overline{[\Pi_1]}$, then $s \in G_\Pi(D) \Leftrightarrow s \in W_{\Pi^n}(D)$ and $s \not \in G_{1_{\Pi_1}}(D)$. Reasoning as above we conclude that $s \in W_{\Pi^n}(D) \Leftrightarrow s \in W_{\Pi^n}(D^t)$. Furthermore, by the induction hypothesis with respect to $\Pi_1$, we get that $s \in G_{1_{\Pi_1}}(D) \Leftrightarrow s \in G_{1_{\Pi_1}}(D^t)$.
        \item   If \ $\Pi = \bigwedge[\Pi_1, \Pi_2]$, then $s \in G_\Pi(D) \Leftrightarrow s \in G_{1_{\Pi_1}}(D)$ and $s \in G_{2_{\Pi_2}}(D) \Leftrightarrow$ (by the induction hypothesis) $s \in G_{1_{\Pi_1}}(D^t)$ and $s \in G_{2_{\Pi_2}}(D^t) \Leftrightarrow s \in G_\Pi(D^t)$.
        \item   If \ $\Pi = \mathbf{X}[\Pi_1]$, then:
        \begin{itemize}
            \item[$(\Rightarrow)$]  Suppose that $s \in G_\Pi(D)$; this is a result of either the first or the second rule of $\Pi$. If it is due to the first rule, then $s \in G_{1_{\Pi_1}}(D)$ and $D$ does not contain a ground fact of the form $R(s, u)$, for any constant $u$. If it is due to the second rule, $D$ contains a ground fact $R(s, u)$, for some constant $u$, and $u \in G_{1_{\Pi_1}}(D)$.

            In the former case, the induction hypothesis implies that $s \in G_{1_{\Pi_1}}(D^t)$. Moreover, by construction $D^t$ contains the ground fact $R(s, s)$. Hence, $s \in G_\Pi(D^t)$ because of the second rule of $\Pi$.

            In the latter case, the induction hypothesis implies that $u \in G_{1_{\Pi_1}}(D^t)$. Taking into account that $D^t$ contains $R(s, u)$, we conclude that $s \in G_\Pi(D^t)$ because of the second rule of $\Pi$.
            \item[$(\Leftarrow)$]   Suppose that $s \in G_\Pi(D^t)$. Let us assume for a moment that $s$ appears in $G_\Pi(D^t)$ due to an application of
the first rule of $\Pi$. This would imply that $s \not \in A_\Pi(D^t)$. But this is absurd because $R^t$ is total by
construction (i.e., $\forall s \exists u R(s, u)$) meaning that $s \in A_\Pi(D^t)$. This shows that when evaluating
$\Pi$ on ``total" databases, such as $D^t$, the first rule of $\Pi$ is redundant. Hence, $s$ must appear in
$G_\Pi(D^t)$ as a result of an application of the second rule of $\Pi$. This means that $D^t$ contains a ground fact of
the form $R(s, u)$, for some constant $u$ (possibly $s = u$), and $u \in G_{1_{\Pi_1}}(D^t)$. The induction hypothesis
gives that $u \in G_{1_{\Pi_1}}(D)$ $(\ast)$. If $D$ contains the ground fact $R(s, u)$, then $s \in G_\Pi(D)$ due to
the second rule of $\Pi$. If however $D$ does not contain the ground fact $R(s, u)$, then by the definition of $R^t$ we
deduce that: (a) $D$ contains no ground fact of the form $R(s, v)$, meaning that $s \not \in A_\Pi(D)$ $(\ast \ast)$
and (b) the ground fact in $D^t$ is actually $R(s, s)$, i.e., $s = u$, which, in view of $(\ast)$, means that $s \in
G_{1_{\Pi_1}}(D)$ $(\ast \ast \ast)$. By $(\ast \ast)$ and $(\ast \ast \ast)$ we conclude that $s \in G_\Pi(D)$ due to
the first rule of $\Pi$.
        \end{itemize}
        \item   If \ $\Pi = \bigcup[\Pi_1, \Pi_2]$, then:
        \begin{itemize}
            \item[$(\Rightarrow)$]  Suppose that $s \in G_\Pi(D)$; from the rules of the program $\Pi$ we see that there is a $s_i$ (possibly $s_i = s$)
such that $s_i \in G_{2_{\Pi_2}}(D)$. In addition, there exists a sequence $s_0 = s, s_1, \ldots, s_i$ such that $D$
contains the ground facts $R(s_r, s_{r+1})$ and $s_r \in G_{1_{\Pi_1}}(D)$ ($0 \leq r < i$). By construction $D^t$ also
contains the ground facts $R(s_r, s_{r+1})$ ($0 \leq r < i$). Further, the induction hypothesis implies that $s_i \in
G_{2_{\Pi_2}}(D^t)$ and $s_r \in G_{1_{\Pi_1}}(D^t)$ ($0 \leq r < i$). Consequently, by successive applications of the
second rule, we conclude that $s \in G_\Pi(D^t)$.

            \item[$(\Leftarrow)$]   Suppose that $s \in G_\Pi(D^t)$. Consider a \emph{minimal} sequence $s_0 = s, s_1, \ldots, s_i$ (possibly $s_i = s$)
such that $D^t$ contains the ground facts $R(s_r, s_{r+1})$, $s_r \in G_{1_{\Pi_1}}(D^t)$ and $s_r \not \in
G_{2_{\Pi_2}}(D^t)$ ($0 \leq r < i$) and $s_i \in G_{2_{\Pi_2}}(D^t)$. Database $D$ also contains the facts $R(s_r,
s_{r+1})$ ($0 \leq r < i$). For suppose to the contrary that $D$ does not contain $R(s_k, s_{k+1})$, for some $k, \ 0
\leq k < i$. This means that $s_k = s_{k+1} = \ldots = s_i$ (recall (\ref{Eq: Total Closure (STD)})), which in turn
implies that $s_i \not \in G_{2_{\Pi_2}}(D^t)$, i.e, a contradiction. Thus, we have established that $D$ also contains
the facts $R(s_r, s_{r+1})$ ($0 \leq r < i$). Now, the induction hypothesis implies that $s \in G_{2_{\Pi_2}}(D)$ and
$s_r \in G_{1_{\Pi_1}}(D)$ ($0 \leq r < i$). Consequently, by successive applications of the second rule, we conclude
that $s \in G_\Pi(D)$.
        \end{itemize}
        \item   If \ $\Pi = \widetilde{\bigcup}[\Pi_1, \Pi_2]$, then let $G_\Pi (D, n)$ and $G_\Pi (D^t, n)$ be the sets of ground facts of $G$ that have been computed during the first $n$ rounds of the evaluation of the last stratum of $\Pi$ on $D$ and $D^t$, respectively. We shall prove that $s \in G_\Pi (D, n) \Leftrightarrow s \in G_\Pi (D^t, n)$ using induction on the number of rounds $n$.
        \begin{itemize}
            \item   ($\Rightarrow$) Let $s \in G_\Pi (D, 1)$; $s$ appears due to one of the first three rules of $\Pi$. If it is due to the first rule: $G(x) \longleftarrow G_1 (x), G_2 (x)$, then $s \in G_{1_{\Pi_1}}(D) \cap G_{2_{\Pi_2}}(D)$ and the induction hypothesis pertaining to $\Pi_1$ and $\Pi_2$ gives that $s \in G_{1_{\Pi_1}}(D^t) \cap G_{2_{\Pi_2}}(D^t)$, which immediately implies that $s \in G_\Pi(D^t, 1)$. If it is due to the second rule: $G(x) \longleftarrow G_2 (x), \neg A(x)$, then $s \in G_{2_{\Pi_2}}(D)$ and $D$ does not contain a ground fact of the form $R(s, u)$, for any constant $u$. The induction hypothesis with respect to $\Pi_2$ implies that $s \in G_{2_{\Pi_2}}(D^t)$. Moreover, by construction $D^t$ contains the ground fact $R(s, s)$. Then, by the fifth rule of $\Pi$: $B(x, y) \longleftarrow G_2(x), R(x, y), G_2(y)$, $(s, s) \in B_\Pi(D^t)$ and, consequently, by the third rule $s \in G_\Pi(D^t, 1)$. If it is due to the third rule: $G(x) \longleftarrow B(x, x)$, then $(s, s) \in B_\Pi(D)$. This means that $D$ contains a sequence of ground facts $R(s_0, s_1), R(s_1, s_2)$, \ldots, $R(s_k, s_{k+1})$ with $s_r \in G_{2_{\Pi_2}}(D), \ 0 \leq r \leq k+1$, and $s_0 = s_{k+1} = s$.  Using the induction hypothesis pertaining to $\Pi_2$ we obtain $s_r \in G_{2_{\Pi_2}}(D^t), \ 0 \leq r \leq k+1$. Further, by construction $D^t$ contains all the facts of $D$ and, therefore, $(s, s) \in B_\Pi(D^t)$. Finally, by the third rule we conclude that $s \in G_\Pi(D^t, 1)$.

                    ($\Leftarrow$) Let $s \in G_\Pi (D^t, 1)$; $s$ appears either due to the first or due to the third rule of $\Pi$. The totality of $R^t$ precludes the use of the second rule. If it is due to the first rule, a trivial invocation of the induction hypothesis pertaining to $\Pi_1$ and $\Pi_2$ gives that $s \in G_\Pi(D, 1)$. If it is due to the third rule, then $(s, s) \in B_\Pi(D^t)$ and $s \in G_{2_{\Pi_2}}(D^t)$. We distinguish two cases, depending on whether $D^t$ contains the ground fact $R(s, s)$ or not. Let us first consider the case where $R(s, s)$ is in $D^t$. If $R(s, s)$ is also in $D$, then, of course, $(s, s) \in B_\Pi(D)$ and, consequently, $s \in G_\Pi(D, 1)$. So, let us assume that $D$ does not contain $R(s, s)$. This means that $D$ contains no ground fact of the form $R(s, u)$, for any $u$, or, in other words, that $s \not \in A_\Pi(D)$. Then, if we apply the second rule of $\Pi$, using the induction hypothesis to derive that $s \in G_{2_{\Pi_2}}(D)$, we conclude that $s \in G_\Pi(D, 1)$. Let us now consider the case where $R(s, s)$ is not in $D^t$. This means that $D^t$ contains a sequence of ground facts $R(s_0, s_1), R(s_1, s_2)$, \ldots, $R(s_k, s_{k+1})$ with $s_r \in G_{2_{\Pi_2}}(D^t), \ 0 \leq r \leq k+1$, and $s_0 = s_{k+1} = s$. Without loss of generality we may assume that this sequence does not contain any fact of the form $R(u, u)$\footnote{To see why, let us suppose that it contains the fact $R(u, u)$. This means that the aforementioned sequence is $R(s_0, s_1), R(s_1, s_2)$, \ldots, $R(s_l, u), R(u, u), R(u, s_{l+3}), R(s_{l+3}, s_{l+4})$, \ldots, $R(s_k, s_{k+1})$. But then simply consider the sequence $R(s_0, s_1), R(s_1, s_2)$, \ldots, $R(s_l, u), R(u, s_{l+3}), R(s_{l+3}, s_{l+4})$, \ldots, $R(s_k, s_{k+1})$ that also gives rise to $(s, s) \in B_\Pi(D^t)$ without containing $R(u, u)$.}. $D$ also contains the facts $R(s_0, s_1), R(s_1, s_2)$, \ldots, $R(s_k, s_{k+1})$; for suppose to the contrary that one of these facts is not present in $D$. But then this ``missing" fact must be of the form $R(u, u)$, which is absurd. Hence, using the induction hypothesis to derive that $s_r \in G_{2_{\Pi_2}}(D), \ 0 \leq r \leq k+1$, we deduce that $(s, s) \in B_\Pi(D)$, and, consequently, that $s \in G_\Pi(D, 1)$.
            \item   We show now that the claim holds for $n+1$, assuming that it holds for $n$.

                    ($\Rightarrow$) Suppose that $s$ first appeared in $G_\Pi (D, n+1)$ during round $n+1$. This could have happened due to one of the first four rules of $\Pi$. In case one of the first three rules is used, then by reasoning as above, we conclude that $s \in G_\Pi(D^t, n+1)$. So, let us suppose that the fourth rule: $G(x) \longleftarrow G_2 (x), R(x, y), G(y)$ is used. This implies that $s \in G_{2_{\Pi_2}}(D)$ and that there exists a $s_1$ such that $R(s, s_1)$ and $s_1 \in G_\Pi(D, n)$. By construction $D^t$ also contains $R(s, s_1)$. Moreover, the induction hypothesis with respect to the number of rounds gives that $s_1 \in G_\Pi(D^t, n)$ and the induction hypothesis with respect to $\Pi_2$ gives that $s \in G_{2_{\Pi_2}}(D^t)$. Thus, by the fourth rule we derive that $s \in G_\Pi(D^t, n+1)$.

                    ($\Leftarrow$) Suppose now that $s$ first appeared in $G_\Pi (D^t, n+1)$ during round $n+1$. This could have happened due to one of the first four rules of $\Pi$. In case one of the first three rules is used, then by reasoning as before, we obtain that $s \in G_\Pi(D, n+1)$. So, let us suppose that the fourth rule: $G(x) \longleftarrow G_2 (x), R(x, y), G(y)$ is used. This implies that $s \in G_{2_{\Pi_2}}(D^t)$ and that there exists a $s_1$ such that $R(s, s_1)$ and $s_1 \in G_\Pi(D^t, n)$. The fact that $s$ first appeared in $G_\Pi (D^t, n+1)$ during round $n+1$ means that $s \neq s_1$ because if $s = s_1$, then $s$ would belong to $G_\Pi(D^t, n)$. This in turn implies that $D$ contains $R(s, s_1)$. Invoking the induction hypothesis we get that $s_1 \in G_\Pi(D, n)$ and $s \in G_{2_{\Pi_2}}(D)$. Thus, by the fourth rule we derive that $s \in G_\Pi(D, n+1)$.

    The bottom-up evaluation of Datalog programs guarantees that there exists $n_0 \in \mathbb{N}$ such that $G_\Pi (D, n_0) = G_\Pi (D, r)$ for every $r > n_0$, meaning that $G_\Pi (D) = G_\Pi (D, n_0)$. Similarly, $G_\Pi (D^t) = G_\Pi (D^t, n_0)$ and, hence, $G_\Pi (D) = G_\Pi (D^t)$. \hfill$\dashv$
        \end{itemize}
\end{enumerate}

\subsection{Embedding STD to CTL}

    We now complete the proof that CTL has exactly the same expressive power with STD programs. The following result complements that of Section \ref{From CTL to STD} and it proves that there exists an embedding of STD to CTL.

\begin{Theorem}         \label{Thrm: From STD to CTL Soundness & Completeness}
    Let $D$ be a relational database over a Kripke schema and let $\mathcal{K}$ be the corresponding finite Kripke
structure. If $\Pi$ is a STD program and $\varphi$ its corresponding CTL formula $($see Definition \ref{Def: STD to CTL
Translation}$)$, then the following holds:
\begin{equation}
    G_\Pi (D) = \varphi [\mathcal{K}]   \label{Eq: From STD to CTL Soundness & Completeness}
\end{equation}
\end{Theorem}

\noindent\textbf{Proof}\\
    From Theorem \ref{Thrm: STD Query Equivalence under Total Closure} we know that $G_\Pi (D) = G_\Pi (D^t)$. Further, we can show that $G_\Pi (D^t) = \varphi [\mathcal{K}]$ -- the proof is identical to the proof of Theorem \ref{Thrm: From CTL to STD Soundness & Completeness} and is omitted. This completes the proof. \hfill$\dashv$

\section{Stratified Datalog: an efficient fragment}  \label{containment}

In Sections \ref{From CTL to STD} and \ref{From STD to CTL} we established the equivalence of CTL with STD. In this
section we capitalize on this relation by showing that STD is an efficient fragment of stratified Datalog in the sense
that: (a) satisfiability and containment are decidable and (b) query evaluation is linear. The only other fragment of
stratified Datalog known to have ``good" properties is presented in \cite{LMSS93} and \cite{HMSS01} where it is shown
that satisfiability and equivalence are decidable for Datalog programs with stratified negation and unary EDB
predicates.

\subsection{Query Evaluation}

Definitions \ref{Def: CTL to STD Translation} and \ref{Def: STD to CTL Translation} in essence provide algorithms for constructing a STD program which corresponds to a CTL formula and vice versa. Notice that this translation can be carried out efficiently in both directions. This is formalized by Propositions \ref{Prop: Linear Construction of STD Queries} and \ref{Prop: Linear Construction of CTL Formulae}, which, together with Propositions \ref{Prop: From Kripke Structures to DBs} and \ref{Prop: From DBs to Kripke Structures}, suggest an efficient method for performing program evaluation in this fragment. Suppose we are given a database $D$ (with a Kripke schema), a STD program $\Pi$ with goal $G$ and we want to evaluate $G$ on $D$, i.e., to compute $G_\Pi(D)$. This can be done as follows:

\begin{enumerate}
    \item   From $\Pi$ and $D$ construct the corresponding $\varphi$ and $\mathcal{K}$ respectively. This step requires
    $O (|\Pi| + |D|)$ time and results in a formula $\varphi$ of size $O (|\Pi|)$ and a Kripke structure $\mathcal{K}$ of
size $O (|D|)$.

    \item   Apply a model checking algorithm for $\mathcal{K}$ and $\varphi$. The algorithm will compile the truth set $\varphi
[\mathcal{K}]$, i.e., the set of states of $\mathcal{K}$ on which $\varphi$ is true. According to Theorem \ref{Thrm:
From STD to CTL Soundness & Completeness}, $\varphi [\mathcal{K}]$ is exactly the outcome of the evaluation of $G$ on
$D$.
\end{enumerate}

    Taking into account that the model checking algorithms for CTL run in $O(|\mathcal{K}| |\varphi|)$ time (see \cite{VW86}), we derive the following theorem.

\begin{Theorem} \label{Thrm: STD Query Evaluation1}
        Given a STD program $\Pi$ with goal $G$ and a database $D$, evaluating $G$ on $D$ can be done in $O(|D| |\Pi|)$ time. \hfill$\dashv$
\end{Theorem}

    The above result establishes the existence of fragments of stratified Datalog where the problem of query evaluation has linear program and data complexity.

\subsection{Satisfiability}

In the following paragraphs we show that the problem of checking the satisfiability of a STD program is reduced to that of checking the satisfiability of a CTL formula. We start with the following corollary of Theorem \ref{Thrm: CTL Validity} and on which we build later to argue about the satisfiability of STD programs.

\begin{Corollary}       \label{Cor: CTL Satisfiability}
    The satisfiability problem for CTL is EXPTIME--complete.
\end{Corollary}

\begin{Definition} $($Satisfiability for Datalog programs$)$     \label{Def: Datalog Satisfiability}
An IDB predicate $G$ of program $\Pi$ is \emph{satisfiable} if there exists a database $D$, such that $G_\Pi (D) \neq \emptyset$.
\hfill\blackrec
\end{Definition}

\begin{Proposition}     \label{Prop: Equivalence of Satisfiability between CTL formulae and STD queries}
    Let $\Pi$ be a STD program with goal predicate $G$ and let $\varphi$ be the corresponding CTL formula; $\varphi$ is satisfiable iff $G$ is satisfiable.
\end{Proposition}
\noindent\textbf{Proof}\\
$(\Rightarrow)$             Suppose that $\varphi$ is satisfiable; then there exists a Kripke structure $\mathcal{K} =
\ang{W, R, V}$, such that $\mathcal{K}, s \models \varphi$, for some $s \in W$. If $\mathcal{K}$ is finite, then by
Theorem \ref{Thrm: From CTL to STD Soundness & Completeness} we obtain that $s \in G_\Pi (D)$, where $D$ is the
database that corresponds to $\mathcal{K}$. If $\mathcal{K}$ is infinite, then by Theorem \ref{Thrm: CTL BMP} there
exists a finite Kripke structure $\mathcal{K}_f = \ang{W_f, R_f, V_f}$ such that $\mathcal{K}_f, s' \models \varphi$,
for some $s' \in W_f$. Invoking Theorem \ref{Thrm: From CTL to STD Soundness & Completeness} we derive that $s' \in
G_\Pi (D)$, where $D$ is the database that corresponds to $\mathcal{K}_f$. We conclude that in both cases $G$ is
satisfiable.

\noindent$(\Leftarrow)$     Suppose now that $G$ is satisfiable. This means that there exists a database $D = \ang{R,
P_0, \ldots, P_n}$ with domain $W$, such that $G_\Pi(D) \neq \emptyset$. Hence, by Theorem \ref{Thrm: From STD to CTL
Soundness & Completeness} we obtain that $\varphi [\mathcal{K}] \neq \emptyset$, where $\mathcal{K}$ is the finite
Kripke structure that corresponds to $D$. This implies that there exists a state $s \in W$ such that $\mathcal{K}, s
\models \varphi$, i.e., $\varphi$ is satisfiable. \hfill$\dashv$
\\

\noindent Proposition \ref{Prop: Equivalence of Satisfiability between CTL formulae and STD queries} provides proof only for the unary goal predicates. The following proposition deals with the case of the binary $B(x,y)$ predicates.

\begin{Proposition}     \label{Prop: Reducing Satisfiability of B to G}
Let $\Pi$ be a STD program and let $B$ be a binary IDB predicate of $\Pi$; the satisfiability of $B$ is reduced in polynomial time to the satisfiability of a unary goal predicate $G$ of a STD program.
\end{Proposition}
\noindent\textbf{Proof}\\
If $\Pi$ contains a binary IDB predicate $B(x, y)$, then it has a subprogram $\Pi' = \widetilde{\bigcup}[\Pi_1, \Pi_2]$. Let $\varphi$, $\psi_1$ and $\psi_2$ be the CTL formulae corresponding to $\Pi', \Pi_1$ and $\Pi_2$. According to Definition \ref{Def: STD to CTL Translation}, $\varphi \equiv \textbf{E}(\psi_1 \widetilde{\mathbf{U}} \psi_2)$; consider now the CTL formula $\varphi^\star \equiv \varphi \wedge \neg \textbf{E} (\top \mathbf{U} \psi_1)$. Let $\Pi''$ be the STD program corresponding to $\varphi^\star$ and let $G$ be the goal predicate of $\Pi''$. But then $B$ is satisfiable iff $G$ is satisfiable. Finally, it is easy to see that the above reduction takes place in polynomial time. \hfill$\dashv$ \\

\noindent    In order to argue about satisfiability we have to argue about the satisfiability of every IDB predicate. A
STD program may contain one of the following  $A$, $W$, $G_i$ and $B_j$ IDB  predicates. The first two predicates are
trivially satisfiable. For the remaining two predicates Propositions \ref{Prop: Equivalence of Satisfiability between
CTL formulae and STD queries} and \ref{Prop: Reducing Satisfiability of B to G} show that they are satisfiable. Thus,
the following theorem is an immediate consequence of Corollary \ref{Cor: CTL Satisfiability} and Propositions
\ref{Prop: Equivalence of Satisfiability between CTL formulae and STD queries} and \ref{Prop: Reducing Satisfiability
of B to G}.

\begin{Theorem}     \label{Thrm: STD Satisfiability}
    The satisfiability problem for STD programs is EXPTIME--complete.
\end{Theorem}

\subsection{Containment}

Deciding the containment of STD programs can also be reduced to the problem of checking the implication of CTL
formulae. First we give some basic definitions regarding the notion of containment for Datalog programs and CTL
formulae.

\begin{Definition}  $($Containment and equivalence of Datalog queries$)$    \label{Def: Datalog Query Containment}
    Given two Datalog queries $\Pi_1$ and $\Pi_2$ with goal predicates $G_1$ and $G_2$, we say that $\Pi_1$ is \emph{contained} in $\Pi_2$, denoted $\Pi_1 \sqsubseteq \Pi_2$, if and only if for every database $D$, $G_{1_{\Pi_1}} (D) \subseteq G_{2_{\Pi_2}} (D)$. $\Pi_1$ and $\Pi_2$ are \emph{equivalent}, denoted $\Pi_1 \equiv \Pi_2$, if $\Pi_1 \sqsubseteq \Pi_2$ and $\Pi_2 \sqsubseteq \Pi_1$. \hfill\blackrec
\end{Definition}

\noindent A similar notion of containment can also be cast in terms of truth sets of CTL formulae.

\begin{Definition}  $($Containment of CTL formulae$)$       \label{Def: CTL Formula Containment}
    Given two CTL formulae $\varphi_1$ and $\varphi_2$, we say that $\varphi_1$ is \emph{contained} in $\varphi_2$, denoted $\varphi_1 \sqsubseteq \varphi_2$, if and only if for every \emph{finite} Kripke structure $\mathcal{K}$,  $\varphi_1[\mathcal{K}] \subseteq \varphi_2[\mathcal{K}]$. \hfill\blackrec
\end{Definition}

    Suppose we have two CTL formulae $\varphi_1$ and $\varphi_2$; we say that $\varphi_1$ \emph{implies} $\varphi_2$ if for every Kripke structure $\mathcal{K} = \ang{W, R, V}$ and for every $s \in W$, $\mathcal{K}, s \models \varphi_1$ implies that $\mathcal{K}, s \models \varphi_2$. If $\varphi_1$ implies $\varphi_2$, then formula $\varphi_1 \rightarrow \varphi_2$ is valid and vice versa. Hence, we use the notation $\models \varphi_1 \rightarrow \varphi_2$ to assert that $\varphi_1$ implies $\varphi_2$ and $\models_f \varphi_1 \rightarrow \varphi_2$ to assert that $\varphi_1$ implies $\varphi_2$ in finite Kripke structures. The following corollary follows directly from Theorem \ref{Thrm: CTL Validity}.

\begin{Corollary}   $($Implication$)$      \label{Cor: CTL Implication} \

    The problem of deciding whether a CTL formula $\varphi_1$ implies a CTL formula $\varphi_2$ is EXPTIME--complete.
\end{Corollary}

The following proposition states that given two CTL formulae $\varphi_1,\varphi_2$,  in order to check implication $\models \varphi_1 \rightarrow \varphi_2$ it is sufficient to check implication only on finite Kripke structures, that is $\models_f \varphi_1 \rightarrow \varphi_2$, because as we have already said CTL exhibits an important property, namely the \textit{bounded model property} (see Theorem \ref{Thrm: CTL BMP}).

\begin{Proposition}     \label{Prop: Equivalence of Implication and Containment}
    Given two CTL formulae $\varphi_1$ and $\varphi_2$ the following are equivalent$:$\\
    $1.$    $\varphi_1 \sqsubseteq \varphi_2$ \\
    $2.$    $\models \varphi_1 \rightarrow \varphi_2$
\end{Proposition}

\noindent\textbf{Proof}\\
$(1 \Rightarrow 2)$             $\varphi_1 \sqsubseteq \varphi_2$ means that for every finite Kripke structure $\mathcal{K} = \ang{W, R, V}$, $\varphi_1(\mathcal{K}) \subseteq \varphi_2(\mathcal{K})$, which implies that if $s \in \varphi_1 (\mathcal{K})$, then $s \in \varphi_2 (\mathcal{K})$ \ ($s \in W$). Therefore, for every finite Kripke structure $\mathcal{K} = \ang{W, R, V}$ and for every $s \in W$, $\mathcal{K}, s \models \varphi_1$ implies $\mathcal{K}, s \models \varphi_2$, that is $\models_f \varphi_1 \rightarrow \varphi_2$.

It remains to consider the infinite case; we will prove that the next two assertions are equivalent:

(a) $\models_f \varphi_1 \rightarrow \varphi_2$

(b) $\models \varphi_1 \rightarrow \varphi_2$

It is obvious that (b) implies (a). To show that (a) also implies (b) we assume, towards contradiction, that (a) holds
and (b) does not hold. This means that $\varphi_1 \wedge \neg \varphi_2$ is satisfiable, i.e., it has a model
$\mathcal{K}$. $\mathcal{K}$ can not be finite because of (a). It must, therefore, be infinite. But then, from Theorem
\ref{Thrm: CTL BMP} we obtain that $\varphi_1 \wedge \neg \varphi_2$ has a finite model $\mathcal{K}_f$, which is a
contradiction because of (a).

\noindent$(2 \Rightarrow 1)$     $\models \varphi_1 \rightarrow \varphi_2$ means that for every Kripke structure
$\mathcal{K} = \ang{W, R, V}$, $\mathcal{K} \models \varphi_1 \rightarrow \varphi_2$. Consequently, for every $s \in
W$, $\mathcal{K}, s \models \varphi_1$ implies that $\mathcal{K}, s \models \varphi_2$, or in other words,
$\varphi_1(\mathcal{K}) \subseteq \varphi_2(\mathcal{K})$. Thus, $\varphi_1 \sqsubseteq \varphi_2$. \hfill$\dashv$
\\

\noindent The following theorem is a direct consequence of Corollary~\ref{Cor: CTL Implication} and
Proposition~\ref{Prop: Equivalence of Implication and Containment}.

\begin{Theorem}     \label{Thrm: CTL Formula Containment}
    The containment problem for CTL formulae is EXPTIME--complete.
\end{Theorem}

\noindent The next theorem follows directly from Theorems \ref{Thrm: From CTL to STD Soundness & Completeness},
\ref{Thrm: From STD to CTL Soundness & Completeness} and \ref{Thrm: CTL Formula Containment}.

\begin{Theorem}     \label{Thrm: STD Query Containment}
    The containment problem for STD programs is EXPTIME--complete.
\end{Theorem}

\noindent\textbf{Proof}\\
Let $\Pi_1, \Pi_2$ be a STD queries with goal predicates $G_1, G_2$ and let $\varphi_1, \varphi_2$ be the corresponding
CTL formulae. We shall prove that $\Pi_1 \sqsubseteq \Pi_2$ iff $\varphi_1 \sqsubseteq \varphi_2$. \\
\noindent ($\Rightarrow$)   Suppose that $\Pi_1 \sqsubseteq \Pi_2$, but it is not the case that $\varphi_1 \sqsubseteq
\varphi_2$, i.e., there exists a finite Kripke structure $\mathcal{K}'$ such that $\varphi_1[\mathcal{K}'] \not
\subseteq \varphi_2[\mathcal{K}']$. Let $D'$ be the database that corresponds to $\mathcal{K}'$; then by Theorem
\ref{Thrm: From CTL to STD Soundness & Completeness} we get that $G_{1_{\Pi_1}} (D') \not \subseteq G_{2_{\Pi_2}}
(D')$. But this is a contradiction because the fact that $\Pi_1 \sqsubseteq \Pi_2$ implies that for every database $D$,
$G_{1_{\Pi_1}} (D) \subseteq G_{2_{\Pi_2}} (D)$. Thus, it must be the case that $\varphi_1 \sqsubseteq \varphi_2$.

\noindent ($\Leftarrow$) Suppose that $\varphi_1 \sqsubseteq \varphi_2$, but it is not the case that $\Pi_1 \sqsubseteq
\Pi_2$, i.e., there exists a database $D'$ such that $G_{1_{\Pi_1}} (D') \not \subseteq G_{2_{\Pi_2}} (D')$. Let
$\mathcal{K}'$ be the finite Kripke structure that corresponds to $D'$. Theorem \ref{Thrm: From STD to CTL Soundness &
Completeness} implies that $\varphi_1[\mathcal{K}'] \not \subseteq \varphi_2[\mathcal{K}']$, which is a contradiction
because $\varphi_1 \sqsubseteq \varphi_2$ means that for every finite Kripke structure $\mathcal{K}$,
$\varphi_1(\mathcal{K}) \subseteq \varphi_2(\mathcal{K})$. Hence, it must be the case that $\Pi_1 \sqsubseteq \Pi_2$.
\hfill$\dashv$
\\

\noindent  The following theorem is an immediate consequence of Theorem \ref{Thrm: STD Query Containment}.

\begin{Theorem}
    The equivalence problem for STD programs is EXPTIME--complete.
\end{Theorem}

\section{Embedding CTL into Datalog$_{Succ}$}      \label{From CTL to TDS}

This section presents an embedding of CTL into a fragment of Datalog$_{Succ}$ that we call the class of Temporal
Datalog Successor (TDS) programs. When embedding CTL into stratified Datalog (Sections 4 and 5) we considered CTL
formulae to be written in existential normal form. In this section we present another aspect of the relation between
CTL and Datalog. In particular, we consider CTL formulae written in positive normal form (see Section 2) and we give an
embedding into Datalog$_{Succ}$. Datalog$_{Succ}$ extends Datalog by assuming ordered domain. In order to express a CTL
formula written in positive normal form in Datalog$_{Succ}$ we need potentially visit all states of the database. For
instance consider the CTL formula  $\mathbf{A} (\psi_1 \mathbf{\widetilde{U}} \psi_2)$. This can be done by using the
$Succ$ predicate of Datalog$_{Succ}$ to count the number of states of the structure reachable from a given state and
check if this number exceeds the cardinality of the database, denoted by $\mathbf{c}_{max}$.

Since Papadimitriou in \cite{Pap85} proved that Datalog$_{Succ}$ captures polynomial time we expect that there is an
embedding from CTL to Datalog$_{Succ}$. In this work we give the exact translation rules.  Recall that $Succ(X,Y)$
means that $Y$ is the successor of $X$, where $X$ and $Y$ take values from a totally ordered domain. We actually use
the equivalent notation $X+1$ for $Succ$ and we make the assumption that the number of elements in the domain is given
and denoted by $\mathbf{c}_{max}$. Note that we use the conventional semantics of Datalog (we compute $G_{\Pi}(D)$ by
computing least fixed points) which constitutes a contribution relatively to work \cite{GFAA03}.

As already mentioned, to traverse all states of a Kripke structure we need the successor build-in predicate $Succ$.
When traversing a path, the order of states is implicitly given by the succession of states on the path. However, this
is not the case when we want to traverse the children of a certain state. So, for this embedding we have to assume that
the set $W_x$ of all children of a certain state $x$ is totally ordered.

In this paragraph we explain in detail how we use the $Succ$ predicate to traverse the Kripke structure. Kripke
structures in essence are directed labeled graphs. In finite Kripke structures every node has a finite branching
degree. In other words for every $x \in W$ there exist $k$ distinct elements $y_0, \ldots, y_{k-1}$ of $W$ such that
$R(x,y_0), \ldots, R(x, y_{k-1})$, for some $k \in \mathbb{N}$ that depends on $x$. When we give the translation rules
of CTL into Datalog$_{Succ}$ it is necessary to capture the relation between a node and its successors. This can be
achieved by introducing the pairwise disjoint relations $S_0, \ldots, S_{k-1}$ (where $k$ is the maximum branching
degree of $\mathcal{K}$) in the corresponding relational database $D$. These relations serve as a ``refinement'' of the
accessibility relation $R$: $R = \bigcup_{i=0}^{k-1} S_i$. Hence, for every node $x$ with $k$ successors we may write
$S_0(x, y_0), \ldots, S_{k-1}(x, y_{k-1})$, instead of $R(x, y_0), \ldots, R(x, y_{k-1})$, meaning that $y_0, \ldots,
y_{k-1}$ are the $1^{st}, \ldots, k^{th}$ children of $x$, respectively. It is easy to see that we can express $S_i's$
using $R$ and $Succ$.

In this translation, for simplicity reasons, we consider Kripke structures of outdegree at most 2. Such structures can
be described by the two disjoint relations $S_0$ and $S_1$ instead of $R$; $S_0(x,y)$ ($S_1(x,y)$) expresses that $y$
is the first (the second child) of $x$. However, $S_1(x,y)$ may not be defined for every state $x$. Notice that due to
the totality of $R$, $S_0(x,y)$ is total i.e., $\forall x \exists y \ S_0(x, y)$.

\subsection{The class TDS}      \label{TDS}

TDS programs are built-up from: (a) two binary ($S_0, S_1$) and an arbitrary number of unary EDB predicates, and (b)
unary and binary IDB predicates. A unary IDB is taken to be the goal predicate of the program.

\begin{Definition}      \label{Def: TDS} \
\begin{itemize}
    \item   The programs
            {\footnotesize $G(x)  \longleftarrow P_i(x)$},
            {\footnotesize $G(x)  \longleftarrow \neg P_i(x)$}  and
            {\footnotesize  $\left\{\begin{array}{l}
                            G(x)  \longleftarrow W(x) \\
                            \Pi^n \\
                            \end{array}\right.$}
are $\mathrm{TDS}_n$ programs having goal predicate $G$.
    \item   If $\Pi_1$ and $\Pi_2$ are $\mathrm{TDS}_n$ programs with goal predicates $G_1$ and $G_2$ respectively, then $\bigwedge[\Pi_1, \Pi_2]$, $\bigvee[\Pi_1, \Pi_2]$, $\mathbf{X}_\exists[\Pi_1]$, $\mathbf{X}_\forall[\Pi_1]$, $\bigcup_\exists[\Pi_1, \Pi_2]$, $\bigcup_\forall[\Pi_1, \Pi_2]$, $\widetilde{\bigcup}_\exists[\Pi_1, \Pi_2]$ and $\widetilde{\bigcup}_\forall[\Pi_1, \Pi_2]$ are also $\mathrm{TDS}_n$ programs with goal predicate $G$.
    \item   The class $\mathrm{TDS}$ is the union of the $\mathrm{TDS}_n$ subclasses:
    \begin{equation}
        TDS = \bigcup_{n \geq 0} TDS_n
    \end{equation}
\end{itemize}
\end{Definition}

In the translation rules we use the notation $X+1$ for the successor of $X$. The program operators $\bigwedge[\cdot,
\cdot]$, $\bigvee [ \cdot, \cdot ]$, $\mathbf{X}_\exists [ \cdot ]$, $\mathbf{X}_\forall [ \cdot ]$, $\bigcup_\exists [
\cdot, \cdot ]$, $\bigcup_\forall [ \cdot, \cdot ]$, $\widetilde{\bigcup}_\exists [ \cdot, \cdot ]$ and
$\widetilde{\bigcup}_\forall [ \cdot, \cdot ]$, depicted in Figure \ref{Fig: TDS Query Operators}, capture the meaning
of the logical connectives $\wedge, \vee$ and the temporal operators $\mathbf{E} \nextt, \mathbf{A} \nextt, \mathbf{E}
\mathbf{U}, \mathbf{A} \mathbf{U}, \mathbf{E} \widetilde{\mathbf{U}}, \mathbf{A} \widetilde{\mathbf{U}}$, respectively.
$\Pi^n$ is used again as an abbreviation for a set of rules. The IDB predicates $W$ and $B$ have the same meaning as in
the STD programs.  As already stated, we use Datalog with the successor built-in predicate (negation is only applied to
EDB predicates). The successor is only required for formulae of the form $\mathbf{A}(\psi_1 \mathbf{\widetilde{U}
\psi_2})$. In this case, the constant $\mathbf{c}_{max}$ is a natural number greater than or equal to 1.
$\mathbf{c}_{max}$ is equal to the cardinality $|W|$ of the underlying temporal Kripke structure $\mathcal{K}$. The
intuition behind operator $\widetilde{\bigcup}_\forall[\cdot, \cdot]$ is the following. The temporal operator
$\mathbf{A} \widetilde{\mathbf{U}}$ holds on a state $s$ if for any path with initial state $s$ either:\\
(1) it is a finite path, $\psi_2$ holds on all its states and $\psi_1$ holds on its last state, or \\
(2) it is an infinite path and $\psi_2$ holds on all its states.

The first, third and fourth rule capture case (1) and they are similar to the rules of ``until". The rest of the rules
capture case (2). Predicate $C(x, n)$ expresses the fact that all paths that start from state $x$ and are of length
less than or equal to $n$, either are assigned $\psi_2$ on all their states up until there is a state assigned
$\psi_1$, or are assigned $\psi_2$ on all their states. The number $\mathbf{c}_{max}$ denotes the maximum number of
states. $C(x, \mathbf{c}_{max})$ establishes that all paths starting from $x$ belong to either case (1) or case (2)
above. If $C(x, \mathbf{c}_{max})$ holds, then all infinite paths starting from $x$, for which (1) above does not hold,
have all their states assigned $\psi_2$. This is true because on a finite graph all paths of length greater than the
number of its nodes contain a cycle. From the six rules with head $C$, the two last are initialization rules ($x$ may
have one child or two children). The other four assert that given any path $\pi$ of length $n$ starting from $x$
either: (a) $\psi_1\mathbf{\widetilde{U}}\psi_2$ is true on $\pi$, or (b)  $\psi_2$ holds on all $n$ states of $\pi$.

\subsection{Translation rules}

In this section we define an embedding from CTL formulae into TDS programs. This is done via a mapping $\mathbf{h}' =
(h'_f, h'_s)$ such that:

\begin{enumerate}
    \item   $h'_f$ maps CTL formulae into TDS programs; $h'_f(\varphi)$ is a program $\Pi$ with unary goal predicate $G$.
    \item   $h'_s$ maps temporal Kripke structures to relational databases, i.e., $h'_s(\mathcal{K})$ is a database $D$.
    \item   For this mapping the following holds:
\begin{displaymath}
    \varphi [\mathcal{K}] = G_\Pi (D), \textrm{where} \ \Pi = h'_f(\varphi) \ \textrm{and} \ D = h'_s(\mathcal{K})
\end{displaymath}
\end{enumerate}

The exact mapping $h'_f$ of CTL formulae into TDS programs is given below. We use the operators of Figure \ref{Fig: TDS
Query Operators} for succinctness and assume that $\Pi_i$ corresponds to subformula $\psi_i, i = 1, 2$.

\begin{Definition}      \label{Def: CTL to TDS Translation}
    Let $\varphi$ be a CTL formula and let $p_0, \ldots, p_n$ be the atomic propositions appearing in $\varphi$. Then $h'_f (\varphi)$ is the $\mathrm{TDS}_n$ program defined recursively as follows:
      \begin{enumerate}
        \item   If $\varphi \equiv p_i$ or $\varphi \equiv \neg p_i$ or $\varphi \equiv \top$, then $h'_f (\varphi)$ is
                {\footnotesize  $\left\{\begin{array}{l}
                                G(x)  \longleftarrow P_i(x) \\
                                \end{array}\right.$},
                 {\footnotesize  $\left\{\begin{array}{l}
                                G(x)  \longleftarrow \neg P_i(x) \\
                                \end{array}\right.$}
                and
                {\footnotesize  $\left\{\begin{array}{l}
                                G(x)  \longleftarrow W(x) \\
                                \Pi^n \\
                                \end{array}\right.$}
                respectively.
        \item   If \ $\varphi \equiv \psi_1 \wedge \psi_2$ or $\varphi \equiv \psi_1 \vee \psi_2$, then $h'_f (\varphi)$ is $\bigwedge[\Pi_1, \Pi_2]$ and $\bigvee[\Pi_1, \Pi_2]$, respectively.
        \item   If $\varphi \equiv \mathbf{E} \nextt \psi_1$ or $\varphi \equiv \mathbf{A} \nextt \psi_1$ or $\varphi \equiv \mathbf{E}(\psi_1 \mathbf{U} \psi_2)$ or $\varphi \equiv \mathbf{A}(\psi_1 \mathbf{U} \psi_2)$ or $\varphi \equiv \mathbf{E} (\psi_1 \widetilde{\mathbf{U}} \psi_2)$, or $\varphi \equiv \mathbf{A} (\psi_1 \widetilde{\mathbf{U}} \psi_2)$, then $h'_f (\varphi)$ is $\mathbf{X}_\exists[\Pi_1]$, $\mathbf{X}_\forall[\Pi_1]$, $\bigcup_\exists[\Pi_1, \Pi_2]$, $\bigcup_\forall[\Pi_1, \Pi_2]$, $\widetilde{\bigcup}_\exists[\Pi_1, \Pi_2]$ and $\widetilde{\bigcup}_\forall[\Pi_1, \Pi_2]$, respectively. \hfill \blackrec
\end{enumerate}
\end{Definition}

\begin{figure}[!htb]
\centering
\textbf{The query operators of the class TDS} \\
{\scriptsize
\begin{tabular} {l}
\\
\hspace{0.9 cm} $\Pi^n =                        \left\{\begin{array}{l}
                                                W(x)  \longleftarrow S_0(x, y)                                          \\
                                                W(x)  \longleftarrow S_0(y, x)                                          \\
                                                W(x)  \longleftarrow S_1(x, y)                                          \\
                                                W(x)  \longleftarrow S_1(y, x)                                          \\
                                                W(x)  \longleftarrow P_0(x)                                             \\
                                                \dots                                                                   \\
                                                W(x)  \longleftarrow P_n(x)                                             \\
                                                \end{array}\right.$                                                             \\
\hspace{0.07 cm} $\bigwedge[\Pi_1, \Pi_2] =     \left\{\begin{array}{l}
                                                G(x) \longleftarrow G_1(x), G_2(x)                                      \\
                                                \Pi_1                                                                   \\
                                                \Pi_2                                                                   \\
                                                \end{array}\right.$                                                             \\
\hspace{0.07 cm} $\bigvee[\Pi_1, \Pi_2] =       \left\{\begin{array}{l}
                                                G(x) \longleftarrow G_1(x)                                              \\
                                                G(x) \longleftarrow G_2(x)                                              \\
                                                \Pi_1                                                                   \\
                                                \Pi_2                                                                   \\
                                                \end{array}\right.$                                                             \\
\hspace{0.4 cm} $\mathbf{X}_\exists[\Pi_1] =    \left\{\begin{array}{l}
                                                G(x) \longleftarrow S_0(x,y), G_1(y)                                    \\
                                                G(x) \longleftarrow S_1(x,y), G_1(y)                                    \\
                                                \Pi_1                                                                   \\
                                                \end{array}\right.$                                                             \\
\hspace{0.4 cm} $\mathbf{X}_\forall[\Pi_1] =    \left\{\begin{array}{l}
                                                G(x) \longleftarrow S_0(x,y), \neg 2S(x), G_1(y)                        \\
                                                G(x) \longleftarrow S_0(x,y), S_1(x, z), G_1(y), G_1(z)                 \\
                                                2S(x) \longleftarrow S_0(x,y), S_1(x, z)                                \\
                                                \Pi_1                                                                   \\
                                                \end{array}\right.$                                                             \\
$\bigcup_\exists[\Pi_1, \Pi_2] =                \left\{\begin{array}{l}
                                                G(x) \longleftarrow G_2(x)                                              \\
                                                G(x) \longleftarrow G_1(x), S_0(x,y), G(y)                              \\
                                                G(x) \longleftarrow G_1(x), S_1(x,y), G(y)                              \\
                                                \Pi_1                                                                   \\
                                                \Pi_2                                                                   \\
                                                \end{array}\right.$                                                             \\
$\bigcup_\forall[\Pi_1, \Pi_2] =                \left\{\begin{array}{l}
                                                G(x) \longleftarrow G_2(x)                                              \\
                                                G(x) \longleftarrow G_1(x), S_0(x,y), \neg 2S(x), G(y)                  \\
                                                G(x) \longleftarrow G_1(x), S_0(x,y), S_1(x,z), G(y), G(z)              \\
                                                2S(x) \longleftarrow S_0(x,y), S_1(x, z)                                \\
                                                \Pi_1                                                                   \\
                                                \Pi_2                                                                   \\
                                                \end{array}\right.$                                                             \\
$\widetilde{\bigcup}_\exists[\Pi_1, \Pi_2] =    \left\{\begin{array}{l}
                                                G(x) \longleftarrow G_1(x), G_{2}(x)                                    \\
                                                G(x) \longleftarrow B(x,x)                                              \\
                                                G(x) \longleftarrow G_2(x), S_0(x, y), G(y)                             \\
                                                G(x) \longleftarrow G_2(x), S_1(x, y), G(y)                             \\
                                                B(x,y) \longleftarrow G_2(x), S_0(x, y), G_2(y)                         \\
                                                B(x,y) \longleftarrow G_2(x), S_1(x, y), G_2(y)                         \\
                                                B(x,y) \longleftarrow G_2(x), S_0(x, u), B(u,y)                         \\
                                                B(x,y) \longleftarrow G_2(x), S_1(x, u), B(u,y)                         \\
                                                \Pi_1                                                                   \\
                                                \Pi_2                                                                   \\
                                                \end{array}\right.$                                                             \\
$\widetilde{\bigcup}_\forall[\Pi_1, \Pi_2] =    \left\{\begin{array}{l}
                                                G(x) \longleftarrow G_1(x), G_{2}(x)                                    \\
                                                G(x) \longleftarrow C(x, \textbf{c}_{max})                              \\
                                                G(x) \longleftarrow G_2(x), S_0(x, y), \neg 2S(x), G(y)                 \\
                                                G(x) \longleftarrow G_2(x), S_0(x,y), S_1(x,z), G(y), G(z)              \\
                                                C(x, n) \longleftarrow G_2(x), S_0(x,y), \neg 2S(x), C(y,n-1), n \leq \mathbf{c}_{max} \\
                                                C(x, n) \longleftarrow G_2(x), S_0(x,y), S_1(x,z), C(y,n-1),            \\
                                                \hspace{1.6 cm} C(z,n-1), n \leq \mathbf{c}_{max}                       \\
                                                C(x, n) \longleftarrow G_2(x), S_0(x,y), S_1(x,z), G(y), C(z,n-1),      \\
                                                \hspace{1.6 cm} n \leq \mathbf{c}_{max}                                 \\
                                                C(x, n) \longleftarrow G_2(x), S_0(x,y), S_1(x,z), C(y,n-1), G(z),      \\
                                                \hspace{1.6 cm} n \leq \mathbf{c}_{max}                                 \\
                                                C(x, 1) \longleftarrow G_2(x), S_0(x,y), \neg 2S(x), G_2(y)             \\
                                                C(x, 1) \longleftarrow G_2(x), S_0(x,y), S_1(x,z), G_2(y), G_2(z)       \\
                                                2S(x) \longleftarrow S_0(x,y), S_1(x, z)                                \\
                                                \Pi_1                                                                   \\
                                                \Pi_2                                                                   \\
                                                \end{array}\right. $
\end{tabular}
} \caption{\textit{\small{These are the query operators used in the definition of the class TDS. $\Pi_1$ and $\Pi_2$
are $\mathrm{TDS}_n$ programs with goal predicates $G_1$ and $G_2$ respectively. $G, B$ and $C$ are ``fresh" predicate
symbols, i.e., they do not appear in $\Pi_1$ or $\Pi_2$. In contrast, $W$ and $2S$ are the same in all programs.
$\Pi^n$ is a convenient abbreviation of the rules depicted here.}}} \label{Fig: TDS Query Operators}
\end{figure}

\noindent Now we are ready to prove the main result of this section.

\begin{Theorem}     \label{Thrm: From CTL to TDS Soundness & Completeness}
    Let $\mathcal{K}$ be a finite Kripke structure and let $D$ be the corresponding relational database. If $\varphi$ is a CTL formula and $\Pi$ its corresponding TDS program, then the following holds:
\begin{equation}
    \varphi [\mathcal{K}] = G_\Pi (D)   \label{Eq: From CTL to TDS Soundness & Completeness}
\end{equation}
\end{Theorem}

\noindent \textbf{Proof}\\
    The proof of (\ref{Eq: From CTL to TDS Soundness & Completeness}) is carried out by induction on the
structure of the formula $\varphi$. The complete proof is presented in the Appendix.             \hfill$\dashv$

\subsection{Unbounded outdegree}

Our results can be easily extended to any Kripke structure with \emph{bounded} outdegree. It is easy to show that even
if we do not have a structure of bounded degree, we can use the order of the domain to express the universal
quantifier. To do so we need the following built-in predicates:

\hspace{0.3cm}(1) $S_0(x,y)$, which says that $y$ is the first (or leftmost) child of $x$, and

\hspace{0.3cm}(2) $Next(x,y)$, which asserts that $y$ is the next sibling of $x$. \\

    For instance, the translation of $\mathbf{A} (\psi_1 \mathbf{U} \psi_2)$ would be the following:

{\footnotesize
\begin{center}
$\left\{\begin{array}{l}
                            G(x) \longleftarrow G_2(x) \\
                            G(x) \longleftarrow G_1(x), S_0(x,y), G(y), B(y) \\
                            B(x) \longleftarrow W(x), \neg N(x) \\
                            B(x) \longleftarrow Next(x,y), G(y), B(y) \\
                            N(x) \longleftarrow Next(x,y)
\end{array}\right.
$\end{center}
}

    In the above program $W$ is the IDB predicate defined by $\Pi^n$ (see Figure \ref{Fig: TDS Query Operators}) that
asserts that $x$ belongs to the domain of the database.

\section{Conclusions and Future Work}

We may express a CTL formula either by omitting the universal quantifier but allowing negation or by restricting
negation to the propositional atoms only and using the universal quantifier. The former yields an embedding into
stratified Datalog  and the latter  into  Datalog$_{Succ}$. Moreover we identify a fragment of stratified Datalog,
called STD, with the same expressive power as CTL.  For STD all the good properties of CTL can be carried over, as the
translation is linear in the size of the formula and the Datalog program. Thus, we derive new results that prove the
decidability of the satisfiability and query containment problems for STD programs by reducing them to the validity
problem for CTL. We also prove that the query evaluation for STD programs can be done in linear time with respect to
the size of the database and the query.

In this paper we work with finite Kripke structures having a total accessibility relation. Our technique can be applied
to infinite tree structures if we also consider greatest fixed points. The translation goes through as it is with the
only difference that for the negation of the until operator we need to use greatest fixed point semantics (see
\cite{GFAA03} for details). In this case, the proof of the theorems given in this paper is similar except the argument
related to the greatest fixed point semantics which however uses the same intuition. We make this remark because it
gives helpful insight but in the present paper we focus on finite structures because the query languages in databases
are applied on (and hence their semantics is restricted to) finite structures.

In future work we plan to extend our approach to CTL$^\star$ (Full Branching Time Logic) \cite{ES84,EH86}. CTL is a
proper and less expressive fragment of CTL$^\star$. Although we believe that the extension is feasible, having
considered and investigated the problem for a short time, we think that the translation of CTL$^\star$ will introduce
additional non-trivial complications.

\bibliography{TOCL}

\section{Appendix: Proof of Theorem  \ref{Thrm: From CTL to TDS Soundness & Completeness}}

\subsection{Preliminary results}

Before giving the proof we introduce some useful notions. Recall that Kripke structures are, in general, directed
labeled graphs and not necessarily trees. Nonetheless, it is convenient to view them as labeled trees, something that
is achieved by unwinding the Kripke structure from a specific node $s$, which is designated as the root of the
resulting tree. Technically, this can be done by using pairs from $W \times \mathbb{N}$, where $\mathbb{N}$ is the set
of natural numbers, instead of just nodes from $W$.

\begin{Definition}      \label{Def: U-Unwinding}
    Given a Kripke structure $\mathcal{K} = \langle W, R, V \rangle$, suppose that $\ \mathcal{K}, s \models \mathbf{A} (\psi_1 \mathbf{U} \psi_2)$. The $\mathbf{U}$ \textit{unwinding} of $\mathcal{K}$ from $s$, denoted $\mathcal{K}_s^\mathbf{U}$, is the Kripke structure $\langle W', R', V' \rangle$, where:
\begin{enumerate}
    \item   $W'$ is the least subset of $W \times \mathbb{N}$ such that:
            \begin{itemize}
                \item   $(s, 0) \in W'$ and
                \item   if $(s', n) \in W'$, $R(s', t)$ holds, $\mathcal{K}, s' \models \psi_1 \wedge \neg \psi_2$ and $\mathcal{K}, t \models \psi_1 \vee \psi_2$, then $(t, n+1) \in W'$
            \end{itemize}
    \item   $R'((s', n), (t, n+1))$ holds iff $R(s', t)$ holds, and

    \item   $V'(t, n) = V(t)$.  \hfill \blackrec
\end{enumerate}
\end{Definition}

\begin{Definition}      \label{Def: U-Tilde-Unwinding}
    Given a finite Kripke structure $\mathcal{K} = \langle W, R, V \rangle$, suppose that $\mathcal{K}, s \models \mathbf{A} (\psi_1 \mathbf{\widetilde{U}} \psi_2)$. The $\mathbf{\widetilde{U}}$ \textit{unwinding} of $\mathcal{K}$ from $s$, denoted $\mathcal{K}_s^\mathbf{\widetilde{U}}$, is the Kripke structure $\langle W', R', V' \rangle$, where:
\begin{enumerate}
    \item   $W'$ is the least subset of $W \times \mathbb{N}$ such that:
            \begin{itemize}
                \item   $(s, 0) \in W'$ and
                \item   if $(s', n) \in W'$, $n < |W| - 1$, $R(s', t)$ holds, $\mathcal{K}, s' \models \psi_2 \wedge \neg \psi_1$ and $\mathcal{K}, t \models \psi_2$, then $(t, n+1) \in W'$
            \end{itemize}
    \item   $R'((s', n), (t, n+1))$ holds iff $R(s', t)$ holds, and

    \item   $V'(t, n) = V(t)$.  \hfill \blackrec
\end{enumerate}
\end{Definition}

The $\mathbf{U}$ and $\mathbf{\widetilde{U}}$ unwindings of a finite Kripke structure $\mathcal{K}$ are finite labeled
trees. Moreover, if $\mathcal{K}$ has branching degree two, then $\mathbf{U}$ and $\mathbf{\widetilde{U}}$ are finite
binary trees. Let $(s', n)$ be a state of $\mathcal{K}_s^\mathbf{U}$ (or $\mathcal{K}_s^\mathbf{\widetilde{U}}$). If
there exists a state $(t, n+1)$ such that $R'((s', n), (t, n+1))$, then $(s', n)$ is an \textit{internal node} of
$\mathcal{K}_s^\mathbf{U}$ (or $\mathcal{K}_s^\mathbf{\widetilde{U}}$); otherwise $(s', n)$ is a \textit{leaf}.

\begin{figure}[htb]
   \centerline{\includegraphics[width=14.0cm]{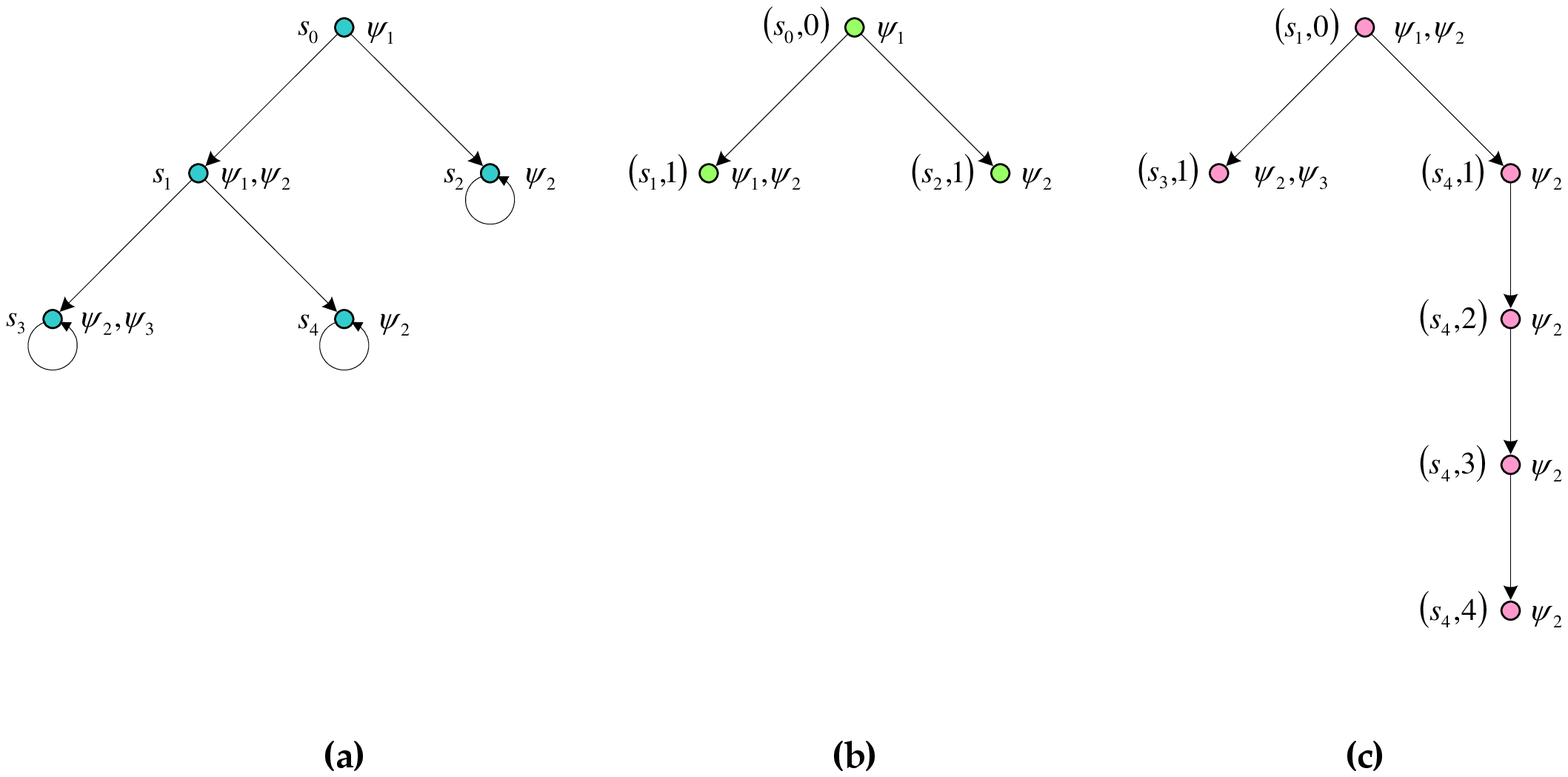}}
   \caption{\textit{\small{Examples of \ $\mathbf{U}$ and $\mathbf{\widetilde{U}}$ unwinding. }}}
   \label{Fig: Unwindings}
\end{figure}

\begin{Example}
    Consider the Kripke structure shown in \emph{Figure \ref{Fig: Unwindings}}$(a)$. The $\mathbf{U}$ unwinding of
$\mathbf{A} (\psi_1 \mathbf{U} \psi_2)$ from $s_0$ is shown in \emph{Figure \ref{Fig: Unwindings}}$(b)$ and the
$\mathbf{\widetilde{U}}$ unwinding of $\mathbf{A} (\psi_3 \mathbf{\widetilde{U}} \psi_2)$ from $s_1$ is depicted in
\emph{Figure \ref{Fig: Unwindings}}$(c)$. Note that in \emph{Figure \ref{Fig: Unwindings}}$(c)$ $(s_4, 3)$ is an
internal node whereas $(s_4, 4)$ is a leaf.  \hfill $\blacktriangle$
\end{Example}

\begin{Proposition}     \label{Prop: U-Unwinding}
    Let $\mathcal{K} = \langle W, R, V \rangle$ be a Kripke structure, let $\mathcal{K}, s \models \mathbf{A} (\psi_1 \mathbf{U} \psi_2)$ and let $\mathcal{K}_s^\mathbf{U}$ be the $\mathbf{U}$ unwinding of $\mathcal{K}$ from $s$. Then the following hold:
\begin{enumerate}
    \item   If $(s', n)$ is a leaf of $\mathcal{K}_s^\mathbf{U}$, then $\mathcal{K}, s' \models \psi_2$.
    \item   If $(s', n)$ is an internal node of $\mathcal{K}_s^\mathbf{U}$, then $\mathcal{K}, s' \models \psi_1 \wedge \neg \psi_2$. \hfill$\dashv$
\end{enumerate}
\end{Proposition}

\begin{Proposition}     \label{Prop: U-Tilde-Unwinding}
    Let $\mathcal{K} = \langle W, R, V \rangle$ be a finite Kripke structure, let $\mathcal{K}, s \models \mathbf{A} (\psi_1 \mathbf{\widetilde{U}} \psi_2)$ and let $\mathcal{K}_s^\mathbf{\widetilde{U}}$ be the $\mathbf{\widetilde{U}}$ unwinding of $\mathcal{K}$ from $s$. Then the following hold:
\begin{enumerate}
    \item   If $(s', n)$ is a leaf of $\mathcal{K}_s^\mathbf{\widetilde{U}}$, then either
            \begin{enumerate}
            \item   $\mathcal{K}, s' \models \psi_1 \wedge \psi_2$, or
            \item   $n = |W|-1$, $\mathcal{K}, s' \models \neg \psi_1 \wedge \psi_2$ and for every child $t$ of $s'$, $\mathcal{K}, t \models \psi_2$.
            \end{enumerate}
    \item   If $(s', n)$ is an internal node of $\mathcal{K}_s^\mathbf{\widetilde{U}}$, then $n < |W| - 1$ and also $\mathcal{K}, s' \models \neg \psi_1 \wedge \psi_2$.
\end{enumerate}
\end{Proposition}

\begin{Proposition}     \label{Prop: Pigeonhole}
    Let $\mathcal{K} = \langle W, R, V \rangle$ be a finite Kripke structure and let $s_0$, \ldots, $s_i$, \ldots, $s_j$, \ldots, $s_n$ be a finite path in $\mathcal{K}$ (or in the corresponding database $D$), where $n \geq |W|$. Then, there exists a state $s$ such that $s_i = s_j = s$.
\end{Proposition}

\subsection{Proof of Theorem        \ref{Thrm: From CTL to TDS Soundness & Completeness}}

We are ready now to prove that (\ref{Eq: From CTL to TDS Soundness & Completeness}) holds by induction on the structure
of formula $\varphi$. To increase the readability of the proof, we use the subscripts in the goal predicates to denote
the corresponding CTL formula. For instance, we write $G_{\mathbf{E} \nextt \psi}$ to denote that $G$ is the goal
predicate of the program corresponding to $\mathbf{E} \nextt \psi$. We consider the two directions separately and begin
by considering the $\Rightarrow$ direction.
\\

\noindent\textbf{Proof ($\Rightarrow$)}
\begin{enumerate}
    \item   If $\varphi \equiv p$ or $\varphi \equiv \neg p$, where $p \in AP$, or $\varphi \equiv \top$, then the corresponding programs are those of
Definition \ref{Def: CTL to TDS Translation}.(1). Trivially, then:
        \begin{itemize}
            \item   ${\mathcal K}, s \models p \Rightarrow p \in V(s) \Rightarrow P(s)$ is a ground fact of $D \Rightarrow s \in G_p(D)$.
            \item   ${\mathcal K}, s \models \neg p \Rightarrow p \not \in V(s) \Rightarrow P(s)$ is not a ground fact of $D \Rightarrow s \in G_{\neg p}(D)$.
            \item   ${\mathcal K}, s \models \top \Rightarrow s \in W \Rightarrow$ (by the totality of $R$) there exists $t \in W$ such that $(s, t) \in S_0 \cup S_1$ $\Rightarrow s \in W_{\Pi^n}(D) \Rightarrow s \in G_\top(D)$.
        \end{itemize}
    \item   If $\varphi\equiv\psi_1\vee\psi_2$ or $\varphi\equiv\psi_1\wedge\psi_2$, then the corresponding programs are shown in Definition \ref{Def: CTL to TDS Translation}.(2). Again, the next hold:
        \begin{itemize}
            \item   ${\mathcal K}, s \models \varphi \Rightarrow {\mathcal K}, s \models \psi_1$ or ${\mathcal K}, s \models \psi_2 \Rightarrow$ (by the induction hypothesis) $s \in G_{\psi_1} (D)$ or $s \in G_{\psi_2} (D) \Rightarrow s\in G_{\psi_1}(D) \cup G_{\psi_2}(D) \Rightarrow s \in G_{\varphi} (D)$.
            \item   ${\mathcal K},s \models \varphi \Rightarrow {\mathcal K},s \models \psi_1$ and ${\mathcal K},s \models \psi_2  \Rightarrow$ (by the induction hypothesis) $s \in G_{\psi_1} (D)$ and $s \in G_{\psi_2} (D) \Rightarrow s \in G_{\psi_1}(D) \cap G_{\psi_2}(D) \Rightarrow s\in G_{\varphi} (D)$.
        \end{itemize}
     \item   If $\varphi \equiv \mathbf{E} \nextt \psi$, then the corresponding program is shown in Definition \ref{Def: CTL to TDS
     Translation}.(3).

          Let us assume that ${\mathcal K}, \pi \models \varphi$ for some path $\pi = s_0, s_1, s_2, \ldots$ \ with initial state $s_0$. We know that either $S_0(s_0, s_1)$ or $S_1(s_0, s_1)$ holds. Now ${\mathcal K}, \pi \models \varphi$ for the path $\pi = s_0, s_1, s_2$, \ldots $\Rightarrow$ ${\mathcal K}, \pi^1 \models \psi$ for the path $\pi^1 = s_1, s_2, \ldots$ $\Rightarrow$ ${\mathcal K}, s_1 \models \psi$ $\Rightarrow$ (by the induction hypothesis) $s_1 \in G_{\psi} (D)$. From $\Pi_\varphi$, by combining $G_{\psi}(s_1)$ with one of $S_0(s_0, s_1)$ or $S_1(s_0, s_1)$, we immediately derive $G_\varphi (s_{0})$ and, thus, $s_{0} \in G_\varphi (D)$.

   \item   If $\varphi \equiv \mathbf{A} \nextt \psi$, then the corresponding program is shown in Definition \ref{Def: CTL to TDS
   Translation}.(3).

Let's assume now that ${\mathcal K},\pi \models \varphi$ for every path $\pi = s_0, s_1, s_2, ... $ with initial state $s_0$. It is convenient to distinguish two cases: \\
    (a) \hspace{0.2 cm} $s_0$ has a left child $s_1^L$, but not a right child. In this case ${\mathcal K}, \pi \models \varphi$ \ for every path $\pi = s_0, s_1, s_2$, \ldots \ with initial state $s_0 \Rightarrow {\mathcal K}, \pi^{1,L} \models \psi$ for every path $\pi^{1,L} = s_1^L, s_2, \ldots$ with initial state $s_1^L$ $\Rightarrow$ ${\mathcal K}, s_1^L \models \psi$ $\Rightarrow$ (by the induction hypothesis) $s_1^L \in G_{\psi} (D)$. Moreover, in this case $S_0 (s_0, s_1^L)$, $\neg 2S(s_0)$ are true and, therefore, evaluation of the second rule of $\Pi_{\varphi}$ gives $G_\varphi (s_0) \longleftarrow S_0 (s_0, s_1^L), \neg 2S(s_0), G_\psi(s_1^L)$ $\Rightarrow$ $s_0 \in G_\varphi (D)$. \\
    (b) \hspace{0.2 cm} $s_0$ has both a left child $s_1^L$ and a right child $s_1^R$. Then ${\mathcal K}, \pi \models \varphi$ for every path $\pi = s_0, s_1, s_2, \ldots$ with initial state $s_0$ $\Rightarrow$ ${\mathcal K}, \pi^{1,L} \models \psi$ \ for every path $\pi^{1,L} = s_1^L, s_2^L, \ldots$ with initial state $s_1^L$ and ${\mathcal K}, \pi^{1,R} \models \psi$ \ for every path $\pi^{1,R} = s_1^R, s_2^R, \ldots$ with initial state $s_1^R$ $\Rightarrow {\mathcal K}, s_1^L \models \psi$ and ${\mathcal K}, s_1^R \models \psi$ $\Rightarrow$ (by the induction hypothesis) $s_1^L, s_1^R \in G_{\psi} (D)$. Moreover, in this case $S_0 (s_0, s_1^L)$, $S_1 (s_0, s_1^R)$ are true and, therefore, evaluation of the third rule of $\Pi_{\varphi}$ gives $G_\varphi (s_0)$ $\longleftarrow$ $S_0(s_0, s_1^L)$, $S_1(s_0, s_1^R)$, $G_\psi(s_1^L)$, $G_\psi(s_1^R)$ $\Rightarrow$ $s_0 \in G_\varphi (D)$.

    \item If $\varphi \equiv \mathbf{E} (\psi_1 \mathbf{U} \psi_2)$, then the corresponding program is this of Definition \ref{Def: CTL to TDS
    Translation}.(3).

       Suppose that ${\mathcal K},\pi \models \varphi$ where the path $\pi$ is $s_0, s_1,$ $s_2, \ldots$ \ . We have to examine two cases: \\
    (a) \hspace{0.2 cm} ${\mathcal K}, \pi \models \psi_2$ for the path $\pi = s_0, s_1, s_2, \ldots$ $\Rightarrow$ ${\mathcal K}, s_0 \models \psi_2$ $\Rightarrow$ (by the induction hypothesis) $s_0 \in G_{\psi_2} (D)$ $\Rightarrow$ $G_{\varphi} (s_0)$ (from the first rule $G_\varphi (x) \longleftarrow G_{\psi_2}(x)$ of $\Pi_{\varphi}$) $\Rightarrow$ $s_{0} \in G_\varphi (D)$. \\
    (b) \hspace{0.2 cm} ${\mathcal K}, \pi^i \models \psi_2$ for the path $\pi^i = s_i, s_{i+1}, s_{i+2}$, \ldots \ and ${\mathcal K}, \pi^j \models \psi_1$ for $\pi^j = s_j, s_{j+1}, s_{j+2},\ldots$ $(0\leq j\leq i-1)$ $\Rightarrow$ ${\mathcal K}, s_i \models \psi_2$ and ${\mathcal K}, s_j \models \psi_1$ $(0\leq j\leq i-1)$ $\Rightarrow$ $s_i \in G_{\psi_2} (D)$ and $s_j \in G_{\psi_1} (D_\mathcal{K})$ $(0\leq j\leq i-1)$ (by the induction hypothesis). We know that for every $r$, $0 \leq r < i$, at least one of $S_0 (s_r, s_{r+1})$ or $S_1(s_r, s_{r+1})$ holds. From the first rule $G_\varphi (x) \longleftarrow G_{\psi_2} (x)$ of $\Pi_{\varphi}$ we derive that $G_{\varphi} (s_i)$. Successive applications of the second ($G_\varphi (s_r)$ $\longleftarrow$ $G_{\psi_1} (s_r)$, $S_0 (s_r,s_{r+1})$, $G_\varphi (s_{r+1})$) and third rule ($G_\varphi (s_r)$ $\longleftarrow$ $G_{\psi_1} (s_r)$, $S_1 (s_r,s_{r+1})$, $G_\varphi (s_{r+1})$) of $\Pi_\varphi$ for every $r$, $0 \leq r < i$, yield $G_\varphi (s_{i-1})$, $G_\varphi (s_{i-2})$, \ldots, $G_\varphi (s_1)$, $G_\varphi (s_0)$. Thus, $s_0 \in G_\varphi (D)$.

\item If $\varphi \equiv \mathbf{A} (\psi_1 \mathbf{U} \psi_2) $, then the corresponding program is this of Definition \ref{Def: CTL to TDS
Translation}.(3).

Let us assume now that ${\mathcal K}, \pi \models \varphi$ for every path $\pi = s_0, s_1, s_2, \ldots$ \ with initial state $s_0$. Consider the $\mathbf{U}$ unwinding $\mathcal{K}_{s_0}^\mathbf{U}$ of ${\mathcal K}$ from $s_0$ and let $(t_0, r)$ be any node of $\mathcal{K}_{s_0}^\mathbf{U}$; we shall prove that $t_0 \in G_\varphi (D)$. This property of $\mathcal{K}_{s_0}^\mathbf{U}$ indeed implies the required result because $(s_0, 0)$ is a node (specifically the root) of $\mathcal{K}_{s_0}^\mathbf{U}$ and, thus, $s_0 \in G_\varphi (D)$. To prove it, let $L_{t_0} = (t_0, r), (t_1, r+1), \ldots, (t_n, r+n)$ be the longest path from $(t_0, r)$ to a leaf $(t_n, r+n)$ of $\mathcal{K}_{s_0}^\mathbf{U}$. We use induction on the length $n$ of the path $L_{t_0}$. \\
    (a) \hspace{0.2 cm} If $n = 0$, then node $(t_0, r)$ itself is a leaf. From Proposition \ref{Prop: U-Unwinding} we know that ${\mathcal K}, t_0 \models \psi_2$ and by the induction hypothesis (pertaining to formula $\psi_2$) we get that $t_0 \in G_{\psi_2} (D)$. Then, from rule $G_\varphi (x) \longleftarrow G_{\psi_2} (x)$ of $\Pi_{\varphi}$, we derive that $G_{\varphi} (t_{0})$. \\
    (b) \hspace{0.2 cm} We show now that the claim holds for paths of length $n+1$, assuming that it holds for paths of length less than or equal to $n$. In this case node $(t_0, r)$ is an internal node of $\mathcal{K}_{s_0}^\mathbf{U}$.  From Proposition \ref{Prop: U-Tilde-Unwinding} we know that ${\mathcal K}, t_0 \models \psi_1$ and by the induction hypothesis (pertaining to formula $\psi_1$) we get that $t_0 \in G_{\psi_1} (D)$. We focus on the case where node $(t_0, r)$ has exactly two successors $(t^L_1, r+1)$ and $(t^R_1, r+1)$ in $\mathcal{K}_{s_0}^\mathbf{U}$ (the case where $(t_0, r)$ has only one successor is easier). Since $L_{t_0}$ has length $n+1$, then both $L_{t^L_1}$ and $L_{t^R_1}$ have length at most $n$. Hence, by the induction hypothesis with respect to the length of the paths $L_{t^L_1}$ and $L_{t^R_1}$, we get that $t_1^L \in G_{\varphi} (D)$ and $t_1^R \in G_{\varphi} (D)$. So $G_{\psi_1} (t_0)$, $S_0(t_0, t_1^L)$, $S_1(t_0, t_1^R)$, $G_{\varphi} (t_1^L)$ and $G_{\varphi} (t_1^R)$ are true and, therefore, the evaluation of the third rule of $\Pi_{\varphi}$ gives that $G_\varphi (t_{0})$ $\longleftarrow$ $G_{\psi_1} (t_0)$, $S_0(t_0, t_1^L)$, $S_1(t_0, t_1^R)$, $G_{\varphi} (t_1^L)$, $G_{\varphi} (t_1^R)$. Thus, $t_0 \in G_\varphi (D)$.

    \item    If $\varphi \equiv \mathbf{E} (\psi_1 \mathbf{\widetilde{U}} \psi_2)$, then the corresponding program is shown in Definition \ref{Def: CTL to TDS
    Translation}.(3).

Suppose that ${\mathcal K}, \pi \models \varphi$ where the path $\pi$ is $s_0, s_1$, $s_2, \ldots$ \ . We must consider two cases: \\
    (a) \hspace{0.2 cm} ${\mathcal K}, \pi^i \models \psi_1 \wedge \psi_2$ for the path $\pi^i = s_i, s_{i+1}$, $s_{i+2}$, \ldots \ and ${\mathcal K}, \pi^j \models \psi_2$ for $\pi^j = s_j, s_{j+1}$, $s_{j+2}$, \ldots \ $(0 \leq j \leq i-1)$ $\Rightarrow$ ${\mathcal K}, s_i \models \psi_1 \wedge \psi_2$ and ${\mathcal K}, s_j \models \psi_2$ $(0\leq j\leq i-1)$ $\Rightarrow$ $s_i \in G_{\psi_1} (D)$ and $s_j \in G_{\psi_2} (D)$ $(0\leq j\leq i)$ (by the induction hypothesis). We know that for every $r$, $0 \leq r < i$, at least one of $S_0 (s_r, s_{r+1})$ or $S_1(s_r, s_{r+1})$ holds. From rule $G_\varphi (x) \longleftarrow G_{\psi_1} (x), G_{\psi_2} (x)$ of $\Pi_{\varphi}$ we derive that $G_{\varphi} (s_{i})$. Successive applications of the other two rules of the program (i.e., $G_\varphi (s_r)$ $\longleftarrow$ $G_{\psi_2} (s_r)$, $S_0 (s_r, s_{r+1})$, $G_\varphi (s_{r+1})$ and $G_\varphi (s_r)$ $\longleftarrow$ $G_{\psi_2} (s_r)$, $S_1 (s_r, s_{r+1})$, $G_\varphi (s_{r+1})$) for every $r$, $0 \leq r < i$, yield $G_\varphi (s_{i-1})$, $G_\varphi (s_{i-2})$, \ldots, $G_\varphi (s_1)$, $G_\varphi (s_0)$. Thus, $s_0 \in G_\varphi (D)$. \\
    (b) \hspace{0.2 cm} ${\mathcal K}, \pi^i \models \psi_2$ for the path $\pi^{i} = s_i, s_{i+1}, s_{i+2}$, \ldots , for every $i \geq 0$. This implies that ${\mathcal K}, s_i \models \psi_2$, for every $i \geq 0$, and (by the induction hypothesis) that $s_i \in G_{\psi_{2}} (D)$, for every $i \geq 0$.  Let $s_0, s_1, s_2, \ldots, s_n$ be an initial segment of $\pi$, where $n = |W|$. From Proposition \ref{Prop: Pigeonhole} we know that in the aforementioned sequence there exists a state $s$ such that $s = s_{k} = s_{l}$, $0 \leq k < l \leq n$. Then $(s_k, s_k) \in B_{\psi_2} (D)$. We know that for every $r$, $0 \leq r < k$, at least one of $S_0 (s_r, s_{r+1})$ or $S_1(s_r, s_{r+1})$ holds. From rule $G_\varphi (x) \longleftarrow B_{\psi_2}(x, x)$ we derive that $G_{\varphi} (s_{k})$. Successive applications of the fourth ($G_\varphi (s_r)$ $\longleftarrow$ $G_{\psi_2} (s_r)$, $S_0 (s_r,s_{r+1})$, $G_\varphi (s_{r+1})$) or the fifth ($G_\varphi (s_r)$ $\longleftarrow$ $G_{\psi_2} (s_r)$, $S_1 (s_r,s_{r+1})$, $G_\varphi (s_{r+1})$) rule of $\Pi_\varphi$ for every $r$, $0 \leq r < k$, yield $G_\varphi (s_{k-1})$, $G_\varphi (s_{k-2})$, \ldots, $G_\varphi (s_1)$, $G_\varphi (s_0)$. Accordingly, $s_0 \in G_\varphi (D)$.

  \item    If $\varphi \equiv \mathbf{A} (\psi_1 \mathbf{\widetilde{U}} \psi_2)$, then the corresponding program is shown in Definition \ref{Def: CTL to TDS
  Translation}.(3).

Let us assume now that ${\mathcal K}, \pi \models \varphi$ for every path $\pi = s_0, s_1, s_2, \ldots$ with initial state $s_0$. Consider the $\mathbf{\widetilde{U}}$ unwinding $\mathcal{K}_{s_0}^\mathbf{\widetilde{U}}$ of ${\mathcal K}$ from $s_0$ and let $(t_0, r)$ be any node of $\mathcal{K}_{s_0}^\mathbf{\widetilde{U}}$. We shall prove that either $G_\varphi (t_0)$ or $C_{\psi_2} (t_0, |W|-r)$ holds. This property of the nodes of $\mathcal{K}_{s_0}^\mathbf{\widetilde{U}}$ ensures that for the root $(s_0, 0)$ it must be the case that $s_0 \in G_\varphi (D)$ (recall the universal program in Section 5). To prove this property, let $L_{t_0} = (t_0, r)$, $(t_1, r+1)$, \ldots, $(t_n, r+n)$ be the longest path from $(t_0, r)$ to a leaf $(t_n, r+n)$ of $\mathcal{K}_{s_0}^\mathbf{\widetilde{U}}$. We are going to use induction on the length $n$ of the path $L_{t_0}$.\\
    (a) \hspace{0.2 cm} If $n = 0$, then node $(t_0, r)$ itself is a leaf. We may assume that node $t_0$ has exactly two successors $t^L_1$ and $t^R_1$ in $D$ because the case where $t_0$ has only one successor can be tackled in the same way. From Proposition \ref{Prop: U-Tilde-Unwinding} we know that there are two cases regarding $t_0$: (1) ${\mathcal K}, t_0 \models \psi_1 \wedge \psi_2$; in this case the induction hypothesis (with respect to $\psi_1$ and $\psi_2$) gives that $t_0 \in G_{\psi_1} (D)$ and $t_0 \in G_{\psi_2} (D)$ and from the first rule of $\Pi_{\varphi}$ we derive that $G_{\varphi} (t_0)$. (2) $r = |W|-1$, ${\mathcal K}, t_0 \models \neg \psi_1 \wedge \psi_2$ and ${\mathcal K}, t^L_1 \models \psi_2$ and ${\mathcal K}, t^R_1 \models \psi_2$. The induction hypothesis with respect to $\psi_2$ gives that $t_0 \in G_{\psi_2} (D)$, $t^L_1 \in G_{\psi_2} (D)$ and $t^R_1 \in G_{\psi_2} (D)$. Using rule $C_{\psi_2}(t_0, 1) \longleftarrow G_{\psi_2}(t_0)$, $S_0(t_0, t^L_1)$, $S_1(t_0, t^R_1)$, $G_{\psi_2}(t^L_1)$, $G_{\psi_2}(t^R_1)$ of $\Pi_{\varphi} ^{\mathbf{A}}$ we conclude that $C_{\psi_2} (t_0, 1)$ holds. \\
    (b) \hspace{0.2 cm} We prove now that the claim holds for paths of length $n+1$, assuming that it holds for paths of length less than or equal to $n$. In this case node $(t_0, r)$ is an internal node of $\mathcal{T}_{s_0}^\mathbf{\widetilde{U}}$. From Proposition \ref{Prop: U-Tilde-Unwinding} we know that $r < |W| - 1$ and ${\mathcal K}, t_0 \models \neg \psi_1 \wedge \psi_2$ and by the induction hypothesis (pertaining to $\psi_2$) we get that $t_0 \in G_{\psi_2} (D)$. We examine the case where node $(t_0, r)$ has exactly two successors $(t^L_1, r+1)$ and $(t^R_1, r+1)$ in $\mathcal{K}_{s_0}^\mathbf{\widetilde{U}}$, where $r + 1 < |W|$. In this case $S_0(t_0, t_1^L)$, $S_1(t_0, t_1^R)$ are true. Since $L_{t_0}$ has length $n+1$, then both $L_{t^L_1}$ and $L_{t^R_1}$ have length at most $n$. Hence, by the induction hypothesis (regarding the path length), we get that $G_\varphi (t_1^L)$ or $C_{\psi_2} (t_1^L, |W|-r-1)$ and $G_\varphi (t_1^R)$ or $C_{\psi_2} (t_1^R, |W|-r-1)$. If $G_\varphi (t_1^L)$ and $G_\varphi (t_1^R)$ are true, then the third rule of $\Pi_{\varphi}$ ($G_\varphi(x)$ $\longleftarrow$ $G_{\psi_2}(x)$, $S_0(x,y)$, $S_1(x,z)$, $G_{\varphi} (y)$, $G_{\varphi} (z)$) implies that $G_\varphi (t_0)$ also holds. If $C_{\psi_2} (t_1^L, |W|-r-1)$ and $C_{\psi_2} (t_1^R, |W|-r-1)$ are true, then using the seventh rule of $\Pi_{\varphi} (C_{\psi_2}(x,n)$ $\longleftarrow$ $G_{\psi_2}(x)$, $S_0(x,y)$, $S_1(x,z)$, $C_{\psi_2} (y,n-1)$, $C_{\psi_2}(z,n-1)$, $n \leq |W|$) we conclude that $C_{\psi_2} (t_0, |W|-r)$ also holds. In the remaining two cases the eighth and ninth rule imply that $C_{\psi_2} (t_0, |W|-r)$.

            We have proved that for the node $(s_0, 0)$ one of $G_\varphi (s_0)$ or $C_{\psi_2} (s_0, |W|)$ holds. If we assume that $C_{\psi_2} (s_0, |W|)$ holds, then the fifth rule of $\Pi_{\varphi}$ implies $G_\varphi (s_0)$. Hence, in any case, $s_0 \in G_\varphi (D)$. \hfill $\dashv$

\end{enumerate}

We complete now the proof of (\ref{Eq: From CTL to TDS Soundness & Completeness}) by examining the opposite direction.
\\
\\

\noindent\textbf{Proof ($\Leftarrow$)}

\begin{enumerate}
    \item   If $\varphi \equiv p$ or $\varphi \equiv \neg p$, where $p \in AP$, or $\varphi \equiv \top$, then the corresponding programs are those of
Definition \ref{Def: CTL to TDS Translation}.(1). Trivially, then:
        \begin{itemize}
            \item   $s\in G_p(D) \Rightarrow P(s)$ is a ground fact of $D \Rightarrow p \in V(s) \Rightarrow {\mathcal K}, s \models p$.
            \item   $s\in G_{\neg p}(D) \Rightarrow P(s)$ is not a ground fact of $D \Rightarrow p \not \in V(s) \Rightarrow {\mathcal K}, s \models \neg p$.
            \item   $s \in G_\top(D) \Rightarrow s \in W_{\Pi^n}(D) \Rightarrow s$ appears in one of $S_0, S_1, P_0, \ldots, P_n \Rightarrow s \in W \Rightarrow {\mathcal K}, s \models \top$.
        \end{itemize}
    \item   If $\varphi\equiv\psi_1\vee\psi_2$ or $\varphi\equiv\psi_1\wedge\psi_2$, then the corresponding programs are shown in Definition \ref{Def: CTL to TDS Translation}.(2). Again, the following hold:
        \begin{itemize}
            \item   $s \in G_{\varphi} (D) \Rightarrow s\in G_{\psi_1}(D) \cup G_{\psi_2}(D_\mathcal{K}) \Rightarrow s \in G_{\psi_1} (D)$ or $s \in G_{\psi_2} (D) \Rightarrow$ (by the induction hypothesis) ${\mathcal K}, s \models\psi_1$ or ${\mathcal K}, s \models\psi_2 \Rightarrow {\mathcal K}, s \models \varphi$.
            \item   $s \in G_{\varphi} (D) \Rightarrow s\in G_{\psi_1}(D) \cap G_{\psi_2}(D_\mathcal{K}) \Rightarrow s \in G_{\psi_1} (D)$ and $s \in G_{\psi_2} (D) \Rightarrow$ (by the induction hypothesis) ${\mathcal K}, s \models\psi_1$ and ${\mathcal K}, s \models\psi_2 \Rightarrow {\mathcal K}, s \models \varphi$.
        \end{itemize}
        \item   If $\varphi \equiv \mathbf{E} \nextt \psi$, then the corresponding program is shown in Definition \ref{Def: CTL to TDS
        Translation}.(3).

Let us assume that $s_0 \in G_\varphi (D)$. From the rules of the program $\Pi_\varphi$ we see that there exists a
$s_1$ such that $G_{\psi} (s_1)$ and also one of $S_0(s_0, s_1)$ or $S_1(s_0, s_1)$ holds. By the induction hypothesis
we get ${\mathcal K}, s_1 \models \psi$. Let $\pi = s_0, s_1, s_2, \ldots$ be any path with initial state $s_0$ and
second state $s_1$. Clearly, then ${\mathcal K}, \pi^1 \models \psi$ for the path $\pi^1 = s_1, s_2, \ldots$ and
${\mathcal K}, \pi \models \varphi$ for the path $\pi = s_0, s_1, s_2, \ldots$.
            \item   If $\varphi \equiv \mathbf{A} \nextt \psi$, then the corresponding program is shown in Definition \ref{Def: CTL to TDS
            Translation}.(3).

Suppose now that $s_0 \in G_\varphi (D)$. It is convenient to distinguish two cases: \\
    (a) \hspace{0.2 cm} $s_0$ has a left successor $s_1^L$ but not a right successor in $D$; in this case $S_0(s_0, s_1^L)$ and $\neg 2S(s_0)$ are true. From the second rule of $\Pi_\varphi$ we see that $G_{\psi} (s_1^L)$ holds. By the induction hypothesis we get ${\mathcal K}, s_1^L \models \psi$. Let $\pi = s_0, s_1, s_2, \ldots$ be an arbitrary path with initial state $s_0$. The fact that $s_0$ has a left successor $s_1^L$ but not a right successor implies that $s_1^L$ is the second state of every such path. Suppose that there exists a path $\pi = s_0, s_1^L, s_2, \ldots$ with initial state $s_0$ such that ${\mathcal K}, \pi \not \models \varphi$. Trivially then ${\mathcal K}, \pi^1 \not \models \psi$, where $\pi^1 = s_1^L, s_2, \ldots$, which in turn implies that ${\mathcal K}, s_1^L \not \models \psi$, which is false. \\
    (b) \hspace{0.2 cm} $s_0$ has both a left successor $s_1^L$ and a right successor $s_1^R$ in $D$; in this case $S_0(s_0, s_1^L)$ and $S_1(s_0, s_1^R)$ are true. From the third rule of $\Pi_\varphi$ we see that $G_{\psi} (s_1^L)$ and $G_{\psi} (s_1^R)$ hold. By the induction hypothesis we get ${\mathcal K}, s_1^L \models \psi$ and ${\mathcal K}, s_1^R \models \psi$. Let $\pi = s_0, s_1, s_2, \ldots$ be an arbitrary path with initial state $s_0$. The fact that $s_0$ has both a left successor $s_1^L$ and a right successor $s_1^R$ implies that either $s_1^L$ or $s_1^R$ is the second state of every such path. Suppose that there exists a path $\pi = s_0, s_1, s_2, \ldots$ with initial state $s_0$ such that ${\mathcal K}, \pi \not \models \varphi$. If that were the case, then ${\mathcal K}, \pi^1 \not \models \psi$, where $\pi^1 = s_1, s_2, \ldots$ \ . But $s_1 = s_1^L$ or $s_1 = s_1^R$, which means that ${\mathcal K}, s_1^L \not \models \psi$ or ${\mathcal K}, s_1^R \not \models \psi$, either of which contradicts the induction hypothesis.

        \item   If $\varphi \equiv \mathbf{E} (\psi_1 \mathbf{U} \psi_2)$, then the corresponding program is this of Definition \ref{Def: CTL to TDS
        Translation}.(3).

Suppose that $s_0 \in G_\varphi (D)$. From the rules of the program $\Pi_\varphi$ we see that there exists a $s_i$
(possibly $s_i = s_0$) such that $G_{\psi_2} (s_i)$ holds. Further, there exists a sequence of states $s_0, s_1,
\ldots, s_i$ such that for every $r$ ($0 \leq r < i$) $G_{\psi_1} (s_r)$ and at least one of $S_0 (s_r, s_{r+1})$ or
$S_1(s_r, s_{r+1})$ is true. By the induction hypothesis we get ${\mathcal K}, s_i \models \psi_2$ and ${\mathcal K},
s_j \models \psi_1$ $(0\leq j\leq i-1)$. Let $\pi = s_0, s_1, s_2, \ldots, s_i, \ldots$ be any path with initial
segment $s_0, s_1, \ldots, s_i$; then ${\mathcal K}, \pi^i \models \psi_2$ and ${\mathcal K}, \pi^j \models \psi_1$ $(0
\leq j \leq i-1)$, i.e., ${\mathcal K}, \pi \models \varphi$.

            \item If $\varphi \equiv \mathbf{A} (\psi_1 \mathbf{U} \psi_2)$ then the corresponding program is this of Definition \ref{Def: CTL to TDS
            Translation}.(3).

Let us assume now that $s_0 \in G_\varphi (D)$. Let us define $G_\varphi (D, n)$ to be the set of ground facts for $G_\varphi$ that have been computed in the first $n$ rounds of the evaluation of program $\Pi_\varphi$. For more details in the bottom-up evaluation of Datalog programs see \cite{Ull88}. We shall prove that for every $t \in G_\varphi (D, n)$, ${\mathcal K},\pi \models \varphi$ for every path $\pi = t_0, t_1, t_2, \ldots$ \ with initial state $t_0 = t$. We use induction on the number of rounds $n$. \\
    (a) \hspace{0.2 cm} If $n = 1$, then $t$ appears in $G_\varphi (D)$ due to the first rule of $\Pi_\varphi$, i.e., $t \in G_{\psi_2} (D)$. By the induction hypothesis with respect to $\psi_2$ we get that ${\mathcal K}, t \models \psi_2$, which trivially implies that ${\mathcal K}, \pi \models \varphi$ for every path $\pi = t_0, t_1, t_2, \ldots$ \ with initial state $t_0 = t$. \\
    (b) \hspace{0.2 cm} We show now that the claim holds for $n+1$, assuming that it holds for $n$. We examine only the case where node $t$ has exactly two successors $t^L$ and $t^R$, since the case where $t$ has only one successor is identical. Without loss of generality we may assume that $t$ first appeared in $G_\varphi (D, n+1)$ during round $n+1$. This must have happened due to the third rule of $\Pi_\varphi$: $G_\varphi(x)$ $\longleftarrow$ $G_{\psi_1}(x)$, $S_0(x,y)$, $S_1(x,z)$, $G_\varphi (y)$, $G_\varphi (z)$. This implies that $t \in G_{\psi_1} (D)$ and both $t^L$ and $t^R$ belong to $G_\varphi (D, n)$. Hence, by invoking the induction hypothesis with respect to $\psi_1$ we get that ${\mathcal K}, t \models \psi_1$, and by the induction hypothesis with respect to the number of rounds we get that ${\mathcal K}, \pi^{1,L} \models \varphi$ for every path $\pi^{1,L} = t_1^L, t_2^L, \ldots$ \ with initial state $t_1^L = t^L$ and ${\mathcal K}, \pi^{1,R} \models \varphi$ for every path $\pi^{1,R} = t_1^R, t_2^R, \ldots$ \ with initial state $t_1^R = t^R$. By combining all these, we conclude that ${\mathcal K}, \pi \models \varphi$ for every path $\pi = t_0, t_1, t_2, \ldots$ \ with initial state $t_0 = t$.

    Note that the bottom-up evaluation of Datalog programs guarantees that there exists $n \in \mathbb{N}$ such that $G_\varphi (D, n) = G_\varphi (D, r)$ for every $r > n$, i.e., $G_\varphi (D) = G_\varphi (D, n)$.

        \item    If $\varphi \equiv \mathbf{E} ( \psi_1 \mathbf{\widetilde{U}} \psi_2)$, then the corresponding program is shown in Definition \ref{Def: CTL to TDS
        Translation}.(3).

Let us assume that $s_0 \in G_\varphi (D)$. Let us define $G_\varphi (D, n)$ to be the set of ground facts for $G_\varphi$ that have been computed in the first $n$ rounds of the evaluation of program $\Pi_\varphi$. We shall prove that for every $t \in G_\varphi (D, n)$, there exists a path $\pi = t_0, t_1, t_2, \ldots$ \ with initial state $t_0 = t$, such that ${\mathcal K}, \pi \models \varphi$. We use induction on the number of rounds $n$. \\
    (a) \hspace{0.2 cm} If $n = 1$, then $t$ appears in $G_\varphi (D)$ due to either the first rule, i.e., $t \in G_{\psi_1} (D) \cap G_{\psi_2} (D)$, or to the third rule, i.e., $(t, t) \in B_{\psi_2} (D)$. In the first case, the induction hypothesis pertaining to $\psi_1$ and $\psi_2$, implies that ${\mathcal K}, t \models \psi_1 \wedge \psi_2$, which immediately implies that ${\mathcal K}, \pi \models \varphi$ for any path $\pi = t_0, t_1, t_2, \ldots$ with initial state $t_0 = t$. In the second case, there is a finite sequence $t_0, t_1, \ldots, t_k$ of states, such that $t_0 = t_k = t$ and $t_j \in G_{\psi_2} (D)$, \ $0 \leq j \leq k$. Thus, by the induction hypothesis, ${\mathcal K}, t_j \models \psi_2$, \ $0 \leq j \leq k$. Consider the path $\pi = (t_0, t_1, \ldots, t_k)^\omega$; for this path we have ${\mathcal K},\pi \models \varphi$. \\
    (b) \hspace{0.2 cm} We show now that the claim holds for $n+1$, assuming that it holds for $n$. We focus on the case where node $t$ has exactly two successors $t^L$ and $t^R$ (the case where $t$ has only one successor is similar). We may further assume that $t$ first appeared in $G_\varphi (D, n+1)$ during round $n+1$. This can only have occurred because of the fourth or fifth rule of $\Pi_\varphi$. Then $t \in G_{\psi_2} (D)$ and at least one of $t^L$ and $t^R$ belongs to $G_\varphi (D, n)$. Without loss of generality, we assume that $t^L \in G_\varphi (D, n)$. By the induction hypothesis, we know that ${\mathcal K}, t \models \psi_2$ and that there exists a path $\pi^1 = t_1, t_2, \ldots$ \ with initial state $t_1 = t^L$, such that ${\mathcal K}, \pi^1 \models \varphi$. Immediately then we conclude that ${\mathcal K}, \pi \models \varphi$, for the path $\pi = t_0, t_1, t_2, \ldots$ \ with $t_0 = t$.

    Note that the bottom-up evaluation of Datalog programs guarantees that there exists $n \in \mathbb{N}$ such that $G_\varphi (D, n) = G_\varphi (D, r)$ for every $r > n$, i.e., $G_\varphi (D) = G_\varphi (D, n)$.

              \item    If $\varphi \equiv \mathbf{A} (\psi_1 \mathbf{\widetilde{U}} \psi_2)$, where $\psi_1$ and $\psi_2$ are state formulae, then the corresponding program  is shown in Definition \ref{Def: CTL to TDS
              Translation}.(3).

Let us suppose now that $s_0 \in G_\varphi (D)$. Let us define $G_\varphi (D, k)$ to be the set of ground facts for $G_\varphi$ that have been computed after $k$ rounds of the evaluation of program $\Pi_{\varphi}$. We shall prove with simultaneous induction on the number of rounds $k$ two things:\\
(1) \hspace{0.2 cm} Let $s \in G_\varphi (D, k)$ and let $\pi = s_0, s_1, \ldots$ be an arbitrary path with initial state $s_0 = s$; then ${\mathcal K}, \pi \models \psi_1 \mathbf{\widetilde{U}} \psi_2$ \\
(2) \hspace{0.2 cm} Let t be a state such that $C_{\psi_2}(t, k)$ holds (here of course $k \leq |W|$) and let $\varrho = t_0, t_1,$ \ldots,$ t_k, \ldots$ be an arbitrary path with initial state $t_0 = t$; then either ${\mathcal K}, \varrho \models \psi_1 \mathbf{\widetilde{U}} \psi_2$ or ${\mathcal K}, t_j \models \psi_2$, for $0 \leq j \leq k$. \\
    (a) \hspace{0.2 cm} If $k = 1$, then $s$ appears in $G_\varphi (D)$ due to the first rule of $\Pi_{\varphi}$, i.e., $s \in G_{\psi_1} (D) \cap G_{\psi_2} (D)$. Hence, ${\mathcal K}, \pi \models \psi_1 \mathbf{\widetilde{U}} \psi_2$, where $\pi = s_0, s_1, \ldots$ is any path with initial state $s_0 = s$.  We assume of course that $|W| > 1$ because the case where $|W| = 1$, that is the database contains only one element, is trivial. Similarly, for any state t, if $C_{\psi_2}(t, 1)$ holds then $C_{\psi_2}(t, 1)$ can only be derived by the tenth or eleventh rule of $\Pi_{\varphi}$. In any case, these rules imply that for every path $\pi = t_0, t_1, \ldots$ with initial state $t_0 = t$ we have that ${\mathcal K}, t_0 \models \psi_2$ and ${\mathcal K}, t_1 \models \psi_2$.\\
    (b) \hspace{0.2 cm} We show now that the claim holds for $k+1$, assuming that it holds for $k$. We consider the case where states $s$ and $t$ have exactly two successors $s^L, s^R$ and $t^L, t^R$, respectively. \\
        (i) \hspace{0.2 cm} Initially, we shall consider the case where $k+1 \leq |W|$.

    \hspace{0.2 cm} We may assume that $s$ \ first appeared in $G_\varphi (D, k+1)$ during round $k+1$. It is important to stress that in this case, $s$ cannot arise from an application of the fifth rule of $\Pi_{\varphi}$ because the first time this may happen is at round $|W| + 1$. This implies that $s$ must have appeared because of the fourth rule $G_\varphi (x)$ $\longleftarrow$ $G_{\psi_2}(x)$, $S_0(x,y)$, $S_1(x,z)$, $G_{\varphi} (y)$, $G_{\varphi} (z)$. Hence, $s \in G_{\psi_2} (D)$ and both of $s^L$ and $s^R$ belong to $G_\varphi (D, k)$. Then, by the induction hypothesis, we get that ${\mathcal K}, s \models \psi_2$, ${\mathcal K}, \pi^{1,L} \models \varphi$ for every path $\pi^{1,L} = s_1^L, s_2^L, \ldots$ \ with initial state $s_1^L = s^L$ and ${\mathcal K}, \pi^{1,R} \models \varphi$ for every path $\pi^{1,R} = s_1^R, s_2^R, \ldots$ \ with initial state $s_1^R = t^R$. Therefore, ${\mathcal K}, \pi \models \varphi$ \ for every path $\pi = s_0, s_1, s_2, \ldots$ \ with initial state $s_0 = s$.

    \hspace{0.2 cm} Moreover, if $C_{\psi_2}(t,k+1)$ holds, then $G_{\psi_2}(t)$ holds and one of the next must also hold:\\
    - $C_{\psi_2}(t^L, k)$ and $C_{\psi_2}(t^R, k)$, \\
    - $G_\varphi (t^L)$ and $C_{\psi_2}(t^R, k)$, or \\
    - $C_{\psi_2}(t^L, k)$ and $G_\varphi (t^R)$.

    \hspace{0.2 cm} Thus, by the induction hypothesis, we know that:\\
    - ${\mathcal K}, t \models \psi_2$, and \\
    - for every path $\varrho^{1,L} = t_1^L, t_2^L, \ldots, t_{k+1}^L, \ldots$ \ with initial state $t_1^L = t^L$ either ${\mathcal K}, \varrho^{1,L} \models \varphi$ or ${\mathcal K}, t_j^L \models \psi_2$ ($1 \leq j \leq k+1$), and \\
    - for every path $\varrho^{1,R} = t_1^R, t_2^R, \ldots, t_{k+1}^R, \ldots$ \ with initial state $t_1^R = t^R$ either ${\mathcal K}, \varrho^{1,R} \models \varphi$ or ${\mathcal K}, t_j^R \models \psi_2$ ($1 \leq j \leq k+1$).

    \hspace{0.2 cm} Taking all these into account, we conclude that for every path $\varrho$ $= t_0$,$ t_1$, $t_2,$ $\ldots$,$ t_{k+1}$, $\ldots$ \ with initial state $t_0 = t$ either ${\mathcal K}, \varrho \models \varphi$ or ${\mathcal K}, t_j \models \psi_2$ ($0 \leq j \leq k+1$). \\
        (ii) \hspace{0.2 cm} Finally, we examine the case where $k+1 > |W|$.

    \hspace{0.2 cm} In this case, $s$ may belong to $G_\varphi (D, k+1)$ either due to the fourth rule ($G_\varphi (x)$ $\longleftarrow$ $G_{\psi_2}(x)$, $S_0(x,y)$, $S_1(x,z)$, $G_{\varphi} (y)$, $G_{\varphi} (z)$) or due to the fifth rule ($G_\varphi (x) \longleftarrow C_{\psi_2}(x,\mathbf{c}_{max})$). If it is due to the fourth rule, then $s \in G_{\psi_2} (D)$ and both of $s^L$ and $s^R$ belong to $G_\varphi (D, k)$, and the proof proceeds as in case (i) above. So, let us suppose that $s$ occurs due to the fifth rule, i.e., $C_{\psi_2}(s, |W|)$ is true. As we have already proved in case (i), this implies that given any path $\pi = s_0, s_1, s_2, \ldots, s_{|W|}, \ldots$ \ with initial state $s_0 = s$, either ${\mathcal K}, \pi \models \psi_1 \mathbf{\widetilde{U}} \psi_2$ or ${\mathcal K}, s_j \models \psi_2$, for $0 \leq j \leq |W|$. If ${\mathcal K}, \pi \models \varphi$ \ for every path $\pi$ with initial state $s_0 = s$, then we are finished because this is exactly what we have to prove. If however this is not the case, then there must be a path $\varrho = s_0, s_1, s_2, \ldots, s_{|W|}, \ldots$ with initial state $s_0 = s$ such that ${\mathcal K}, \varrho \not \models \psi_1 \mathbf{\widetilde{U}} \psi_2$. But then for the path $\varrho = s_0, s_1, s_2, \ldots, s_{|W|}, \ldots$ we would have that ${\mathcal K}, s_j \models \psi_2$, for $0 \leq j \leq |W|$ $(\ast)$. Now ${\mathcal K}, \varrho \not \models \psi_1 \mathbf{\widetilde{U}} \psi_2$ $\Rightarrow$ ${\mathcal K}, \varrho \models \neg (\psi_1 \mathbf{\widetilde{U}} \psi_2)$ $\Rightarrow$ ${\mathcal K}, \varrho \models \neg \psi_1 \mathbf{U} \neg \psi_2$ $\Rightarrow$ there exists $i \geq 0$ such that ${\mathcal K}, \varrho^i \models \neg \psi_2$ and for every $j, \ 0 \leq j < i$, ${\mathcal K}, \varrho^j \models \neg \psi_1$ $(\ast \ast)$. The fact that ${\mathcal K}, \varrho^{i} \models \neg \psi_2$ immediately implies that ${\mathcal K}, s_{i} \models \neg \psi_2$ $(\ast \ast \ast)$ ($\psi_2$ is a state formula). If $i \leq |W|$, then a contradiction is immediate because we would have ${\mathcal K}, s_{i} \models \neg \psi_2$ and ${\mathcal K}, s_{i} \models \psi_2$ due to $(\ast)$. Hence, it remains to examine the case where $i > |W|$. Let us examine the initial segment $s_0, s_1, s_2$, \ldots, $s_{|W|}, \ldots, s_{i}$ of $\varrho$; from Proposition \ref{Prop: Pigeonhole} we know that in this initial segment there exists a state $s'$ such that $s' = s_l = s_{l'}$, $0 \leq l < l' \leq i$. So we can get the sorter initial segment $t_0, \ldots, t_{l-1}, t_{l}, t_{l+1}, \ldots, t_{i-(l'-l)}$, where $t_r = s_r, \ 0 \leq r \leq l$ and $t_r = s_{r+(l'-l)}, \ l+1 \leq r \leq i-(l'-l)$. Now if $i-(l'-l)$ is less than or equal to $|W|$, we stop; otherwise we keep applying the same technique until we eventually produce an initial segment $t_0, \ldots, t_m$, where $t_0 = s$, $t_m = s_{i}$ and $m \leq |W|$. Consider the path $\sigma =  t_0, \ldots, t_m, \varrho^{i+1}$ (a shortened version of $\varrho$); we know from $(\ast \ast)$ that ${\mathcal K}, \sigma^m \models \neg \psi_2$ and for every $j, \ 0 \leq j < m$, ${\mathcal K}, \sigma^j \models \neg \psi_1$, which implies that ${\mathcal K}, \sigma \not \models \psi_1 \mathbf{\widetilde{U}} \psi_2$. The only other possibility left for $\sigma$ is that ${\mathcal K}, t_j \models \psi_2$, for $0 \leq j \leq |W|$ \ and, thus, ${\mathcal K}, s_i \models \psi_2$, which contradicts $(\ast \ast \ast)$. This concludes the proof that ${\mathcal K}, \pi \models \varphi$ \ for every path $\pi$ with initial state $s_0 = s$. The bottom-up evaluation of Datalog programs guarantees that there exists $n \in \mathbb{N}$ such that $G_\varphi (D, n) = G_\varphi (D, r)$ for every $r > n$, i.e., $G_\varphi (D) = G_\varphi (D, n)$. \hfill $\dashv$
\end{enumerate}

\end{document}